\documentclass[aps,prc,superscriptaddress,showpacs,amsfonts,twocolumn,floatfix]{revtex4}
\usepackage{graphics}
\usepackage{latexsym}
\usepackage{subfigure,wrapfig}
\usepackage{epsfig,color,rotating,amsmath,delarray,array}
\usepackage{makeidx,pifont,float,amssymb}
\definecolor{gray01}{gray}{0.9}
\definecolor{gray02}{gray}{0.8}
\definecolor{gray03}{gray}{0.7}
\definecolor{gray04}{gray}{0.6}
\definecolor{gray05}{gray}{0.5}
\definecolor{gray06}{gray}{0.4}
\definecolor{gray07}{gray}{0.3}
\definecolor{gray08}{gray}{0.2}
\definecolor{gray09}{gray}{0.1}

\begin{document}

\title{Photoproduction of $\boldsymbol\eta$ and $\boldsymbol\eta\,^\prime$ Mesons off Protons}

\newcommand*{\HISKP}{Helmholtz-Institut f\"ur Strahlen- und Kernphysik der Universit\"at Bonn, Germany}
\newcommand*{\GATCHINA}{Petersburg Nuclear Physics Institute, Gatchina, Russia}
\newcommand*{\PI}{Physikalisches Institut, Universit\"at Bonn, Germany}
\newcommand*{\KVI}{KVI, Groningen, Netherlands}
\newcommand*{\FSU}{Department of Physics, Florida State University, Tallahassee, FL, USA}
\newcommand*{\GIESSEN}{II. Physikalisches Institut, Universit\"at Giessen}
\newcommand*{\BASEL}{Physikalisches Institut, Universit\"at Basel, Switzerland}

\author{V.~Crede} \affiliation{\FSU}
\author{A.~McVeigh} \affiliation{\FSU}
\author{A.V.~Anisovich} \affiliation{\HISKP}\affiliation{\GATCHINA}
\author{J.C.S.~Bacelar} \affiliation{\KVI}
\author{R.~Bantes} \affiliation{\HISKP}
\author{O.~Bartholomy} \affiliation{\HISKP}
\author{D.~Bayadilov} \affiliation{\HISKP}\affiliation{\GATCHINA}
\author{R.~Beck} \affiliation{\HISKP}
\author{Y.A.~Beloglazov} \affiliation{\GATCHINA}
\author{R.~Castelijns} \affiliation{\KVI}
\author{A.~Ehmanns} \affiliation{\HISKP}
\author{D.~Elsner} \affiliation{\PI}
\author{K.~Essig} \affiliation{\HISKP}
\author{R.~Ewald} \affiliation{\PI}
\author{I.~Fabry} \affiliation{\HISKP}
\author{M.~Fuchs} \affiliation{\HISKP}
\author{Chr.~Funke} \affiliation{\HISKP}
\author{R.~Gothe} \affiliation{\PI}
\author{R.~Gregor} \affiliation{\GIESSEN}
\author{A.~Gridnev} \affiliation{\GATCHINA}
\author{E.~Gutz} \affiliation{\HISKP}
\author{St.~H\"offgen} \affiliation{\PI}
\author{P.~Hoffmeister} \affiliation{\HISKP}
\author{I.~Horn} \affiliation{\HISKP}
\author{I.~Jaegle} \affiliation{\BASEL}
\author{J.~Junkersfeld} \affiliation{\HISKP}
\author{H.~Kalinowsky} \affiliation{\HISKP}
\author{S.~Kammer} \affiliation{\PI}
\author{Frank~Klein} \affiliation{\PI}
\author{Friedrich~Klein} \affiliation{\PI}
\author{E.~Klempt} \affiliation{\HISKP}
\author{M.~Konrad} \affiliation{\PI}
\author{M.~Kotulla} \affiliation{\BASEL}\affiliation{\GIESSEN}
\author{B.~Krusche} \affiliation{\BASEL}
\author{J.~Langheinrich} \affiliation{\PI}
\author{H.~L\"ohner} \affiliation{\KVI}
\author{I.V.~Lopatin} \affiliation{\GATCHINA}
\author{J.~Lotz} \affiliation{\HISKP}
\author{S.~Lugert} \affiliation{\GIESSEN}
\author{D.~Menze} \affiliation{\PI}
\author{T.~Mertens} \affiliation{\BASEL}
\author{J.G.~Messchendorp} \affiliation{\KVI}\affiliation{\GIESSEN}
\author{V.~Metag} \affiliation{\GIESSEN}
\author{M.~Nanova} \affiliation{\GIESSEN}
\author{V.A.~Nikonov} \affiliation{\HISKP}\affiliation{\GATCHINA}
\author{D.~Novinski} \affiliation{\GATCHINA}
\author{R. Novotny} \affiliation{\GIESSEN}
\author{M.~Ostrick} \affiliation{\PI}
\author{L.M. Pant} \affiliation{\GIESSEN}
\author{H.~van~Pee} \affiliation{\HISKP}
\author{M.~Pfeiffer} \affiliation{\GIESSEN}
\author{A.~Roy} \affiliation{\GIESSEN}
\author{A.V.~Sarantsev} \affiliation{\HISKP}\affiliation{\GATCHINA}
\author{S.~Schadmand} \affiliation{\GIESSEN}
\author{C.~Schmidt} \affiliation{\HISKP}
\author{H.~Schmieden} \affiliation{\PI}
\author{B.~Schoch} \affiliation{\PI}
\author{S.~Shende} \affiliation{\KVI}
\author{V.~Sokhoyan} \affiliation{\HISKP}
\author{N.~Sparks} \affiliation{\FSU}
\author{A.~S{\"u}le} \affiliation{\PI}
\author{V.V.~Sumachev} \affiliation{\GATCHINA}
\author{T.~Szczepanek} \affiliation{\HISKP}
\author{U.~Thoma} \affiliation{\HISKP}
\author{D.~Trnka} \affiliation{\GIESSEN}
\author{R.~Varma} \affiliation{\GIESSEN}
\author{D.~Walther} \affiliation{\HISKP}\affiliation{\PI}
\author{Ch.~Weinheimer} \affiliation{\HISKP}
\author{Ch.~Wendel} \affiliation{\HISKP}
\author{A.~Wilson} \affiliation{\FSU}
\collaboration{The CBELSA/TAPS Collaboration} \noaffiliation

\begin{abstract}
Total and differential cross sections for $\eta$ and $\eta\,^\prime$ 
photoproduction off the proton have been determined with the CBELSA/TAPS 
detector for photon energies between 0.85 and 2.55~GeV. 
The $\eta$~mesons are detected in their two neutral decay modes, 
$\eta\to\gamma\gamma$ and $\eta\to 3\pi^0\to 6\gamma$, and for the first 
time, cover the full angular range in $\rm cos\,\theta_{cm}$ of the 
$\eta$~meson. These new $\eta$~photoproduction data are consistent with 
the earlier CB-ELSA results. The $\eta\,^\prime$ mesons are observed in 
their neutral decay to $\pi^0\pi^0\eta\to 6\gamma$ and also extend the
coverage in angular range.
\end{abstract}

\date{Received: \today / Revised version:}


\maketitle

\section{\label{Section:Introduction}Introduction}
Understanding the structure of the proton and its excited states is one of
the key questions in hadronic physics. Known as the missing-baryon
problem, quark models based on three constituent quark degrees of freedom
predict many more states than have been observed experimentally. Baryon
resonances are broad and widely overlap, especially at higher energies,
imposing challenges on the interpretation of experimental data in terms of
resonance contributions. Without precise data from many decay channels, it 
will be difficult or even impossible to accurately determine the properties 
of well established resonances, or to confirm or rule out the existence of 
weakly established resonances or new, so-far not observed states.

Of particular importance are well-chosen decay channels which can help isolate
contributions from individual excited states and clarify their importance.
Photoproduction of $\eta$ and $\eta\,^\prime$ mesons offers the distinct 
advantage of serving as an {\it isospin filter} for the spectrum of nucleon 
resonances and thus, simplifies data interpretations and theoretical efforts 
to predict the excited states contributing to these reactions. Since the $\eta$ 
and $\eta\,^\prime$ mesons have isospin $I=0$, the $N\eta$ and $N\eta\,^\prime$ 
final states can only originate from intermediate $I=1/2$ nucleon states.

Data on $\eta$ photoproduction off the free proton were obtained and studied at 
many different laboratories over a wide kinematic range~\cite{Krusche:1995nv,
Ajaka:1998zi,Bock:1998rk,Armstrong:1998wg,Thompson:2000by,Renard:2000iv,Dugger:2002ft,
Crede:2003ax,Elsner:2007hm,Denizli:2007tq}. A review of the main data sets and 
a corresponding comparison of their coverage in energy and solid angle can be 
found in~\cite{Bartholomy:2007zz}. Almost all analyses found that the $N(1535)S_{11}$ 
nucleon resonance dominates $\eta$~photoproduction at threshold, though there 
are models which do not need the $N(1535)S_{11}$~resonance to describe threshold
production of $\eta$~mesons~\cite{Denschlag:1998qn}. The $N(1535)S_{11}$~state 
is well-known for its large $N\eta$ coupling, whereas other resonances couple 
only weakly to $N\eta$. Small contributions from the $N(1520)D_{13}$ resonance 
via interference with the $S_{11}$~resonance have been determined from data on 
photon beam asymmetries~\cite{Ajaka:1998zi,Elsner:2007hm} and angular distributions. 
Data from target polarization experiments~\cite{Bock:1998rk} revealed surprising 
effects concerning the phase relations of the $s$- and $d$-wave amplitudes
\cite{Tiator:1998qp}. Despite its four-star assignment by the Particle Data Group 
(PDG)~\cite{Amsler:2008zz}, the role and nature of the $N(1535)S_{11}$ is still 
not well understood. Surprisingly, the $N(1650)S_{11}$ nucleon resonance has the 
same quantum numbers, but shows no strong $N\eta$ coupling. Many different arguments 
have been discussed to explain this observation; the two $S_{11}$ states can have 
appreciable mixing~\cite{Isgur:1977ef}, for instance. The $N(1535)S_{11}$ resonance 
can also be a dynamically-generated state of the $\Sigma\, K-p\eta$ 
system~\cite{Kaiser:1995cy} or more generally, a dynamically generated resonance 
coming from the interaction of the octet of pseudoscalar mesons with the ground-state 
octet of baryons~\cite{Jido:2003cb,Jido:2007sm}.
Recent efforts at Jefferson Laboratory have concentrated on describing the $\gamma 
p\to N^\ast$ transitions by the interaction of the photon with the \text{3-quark} 
core of the resonance including meson-cloud effects in the low $Q^2$ 
region~\cite{JuliaDiaz:2007fa}. The agreement of the model predictions for the 
helicity amplitude $A_{1/2}$ with experimental data is good for some lower-lying 
$N^\star$~states to fair for the $N(1535)S_{11}$ not ruling out alternative explanations.

The importance of contributions from the $N(1650)S_{11}$ resonance to $\eta$ 
photoproduction has been discussed further in conjunction with its photoproduction 
off the neutron. Recently, the neutron data have attracted interest due to the 
observation of a bump-like structure at 1.67~GeV/$c^2$~\cite{Kuznetsov:2004gy,
Jaegle:2008ux}, which has not been seen in the cross section off the proton. 
In~\cite{Anisovich:2008wd}, it has been shown that a strong interference between 
$S_{11}(1535)$, $S_{11}(1650)$, and a non-resonant background can provide a
good description of these data. 

A partial wave analysis (PWA) of recent CB-ELSA data in the framework of the 
Bonn-Gatchina (BnGa) model~\cite{Anisovich:2005tf}, which included data on other 
reactions and from other experiments, found the dominance of three nucleon resonances 
in $\eta$ photoproduction: $N(1535)S_{11}$, $N(1720)P_{13}$, and a proposed new 
state, $N(2070)D_{15}$~\cite{Crede:2003ax}. The large $N\eta$ coupling of the 
$N(1720)P_{13}$ was surprising. Solutions of the $\eta$-MAID model~\cite{Chiang:2002vq} 
in this mass range assign much of the intensity to the $N(1710)P_{11}$ instead. 
Current efforts with regard to the extraction of double-polarization observables
will help shed light on this controversy. The data presented here cover the full 
angular range and within the framework of the Bonn-Gatchina model, are still consistent 
with the dominance of the three nucleon resonances, $N(1535)S_{11}$, $N(1720)P_{13}$, 
and $N(2070)D_{15}$.

Data on $\eta\,^\prime$ photoproduction is scarce. Analyses published before 2005
observed fewer than 300~$\eta\,^\prime$~events~\cite{ABBHHM,AHHM,Plotzke:1998ua}
and an interpretation in terms of resonance contributions was difficult. Data 
from Jefferson Laboratory significantly improved the world database
\cite{Dugger:2005my}. They observed $2\times 10^5~\eta\,^\prime$~events, which
allowed the extraction of differential cross sections. Though more precise than 
previous measurements, the CLAS data are still limited in their angular coverage. 
In the model by Nakayama and Haberzettl (\cite{Dugger:2005my} and ref.~[12] therein),
the $N\eta\,^\prime$~final state couples to $N(1535)S_{11}$ and $N(1710)P_{11}$.
The authors claim the importance of $J=3/2$ states ($N(1940)P_{13}$, $N(1780)D_{13}$, 
$N(2090)D_{13}$) in the process, which are useful to obtain the correct shape of the 
differential cross sections for energies from 1.728~GeV to 1.879~GeV.

In this paper, we present total and differential cross sections for the reactions:
\begin{align}
\gamma p\to p\eta,&\quad\text{ where }\eta\to 2\gamma,\label{ReactionEta1}\\
\gamma p\to p\eta,&\quad\text{ where }\eta\to 3\pi^0\to 6\gamma,\text{ and}\label{ReactionEta2}\\
\gamma p\to p\eta\,^\prime,&\quad\text{ where }\eta\,^\prime\to 2\pi^0\eta\to 6\gamma.\label{ReactionEtap}
\end{align}

\noindent
The data cover an incoming photon energy range up to 2.55~GeV and show the full
angular coverage.
\begin{figure}[pt]
  \epsfig{file=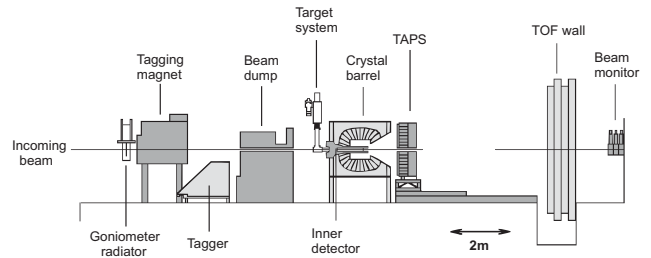,width=.5\textwidth}
    \caption{\label{Figure:Experiment} Experimental setup of CBELSA/TAPS
      in Bonn. The electron beam delivered by the accelerator ELSA enters 
      from the left side.}
\end{figure}

The paper has the following structure. Section~\ref{Section:ExperimentalSetup}
gives a brief introduction to the CBELSA/TAPS experimental setup. The data
reconstruction and selection is discussed in section \ref{Section:DataAnalysis}
and the extraction of differential and total cross sections is described in
section \ref{Section:DeterminationCS}. Experimental results 
are finally presented in section \ref{Section:Results}.

\section{\label{Section:ExperimentalSetup}Experimental Setup}
The experiment was carried out at the electron accelerator facility 
ELSA~\cite{Hillert:2006yb} at the University of Bonn using a combination of 
the Crystal-Barrel~\cite{Aker:1992} and TAPS~\cite{Novotny:1991ht,Gabler:1994ay} 
detectors. The experimental setup is shown in Fig.~\ref{Figure:Experiment}.

Electrons with an energy of 3.175~GeV were extracted from ELSA via
slow (resonant) extraction. The bremsstrahlung-tagger photon-beam
facility at ELSA delivered unpolarized tagged-photon beams in the energy
range from 0.5 to 2.9~GeV by passing the electron beam through a
thin copper radiator with a thickness of $(3/1000)\cdot X_R$ (radiation
length). Electrons are deflected in the field of the tagger dipole magnet
according to their energy loss in the bremsstrahlung process; the remaining 
energy is determined in a tagger detector consisting of 480~scintillating 
fibers above 14~scintillation counters (tagger bars) in a configuration 
with adjacent paddles partially overlapping. The corresponding energy of an
emitted photon is $E_\gamma = E_0 - E_{\rm e^-}$. Electrons not undergoing 
bremsstrahlung are deflected at small angles and guided into a beam dump 
located behind the tagger detectors. The energy resolution is about 2~MeV 
for the high-energy photons and 25~MeV for the low-energy part of the 
bremsstrahlung spectrum.

For the energy calibration of the tagger, a polynomial was
determined in simulations using the measured field map of the bending
magnet and the known positions of the fibers. The calibration was
cross-checked by measurements with the ELSA electron beam at two different 
energies. At 600 and 800~MeV, a low-current beam was guided directly 
into the tagger, while the magnetic field was slowly varied. These results 
provided corrections to the initial polynomial~\cite{PhD:FrankKlein}.

\begin{figure}
  \begin{center}
    \epsfig{file=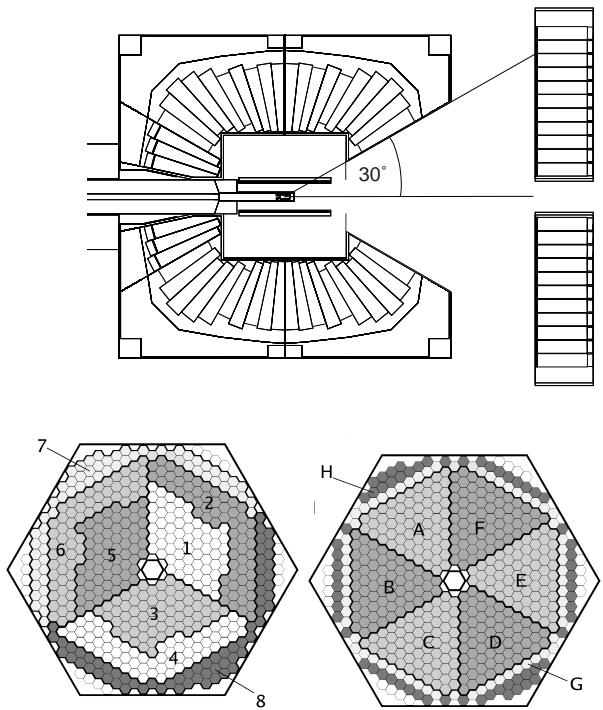,width=0.5\textwidth}
  \end{center}
  \caption{\label{Figure:CB-Luzy-H2} Top: Schematic drawing of the liquid
    hydrogen target, scintillating-fibre detector, Crystal-Barrel and 
    TAPS calorimeters. Bottom: Front view of TAPS; the left side shows the 
    logical segmentation for the LED-low trigger, the right side the logical 
    segmentation for the LED-high trigger (see text for more details).}
\end{figure}

\begin{figure*}
  %\begin{tabular}{cc}
    \epsfig{file=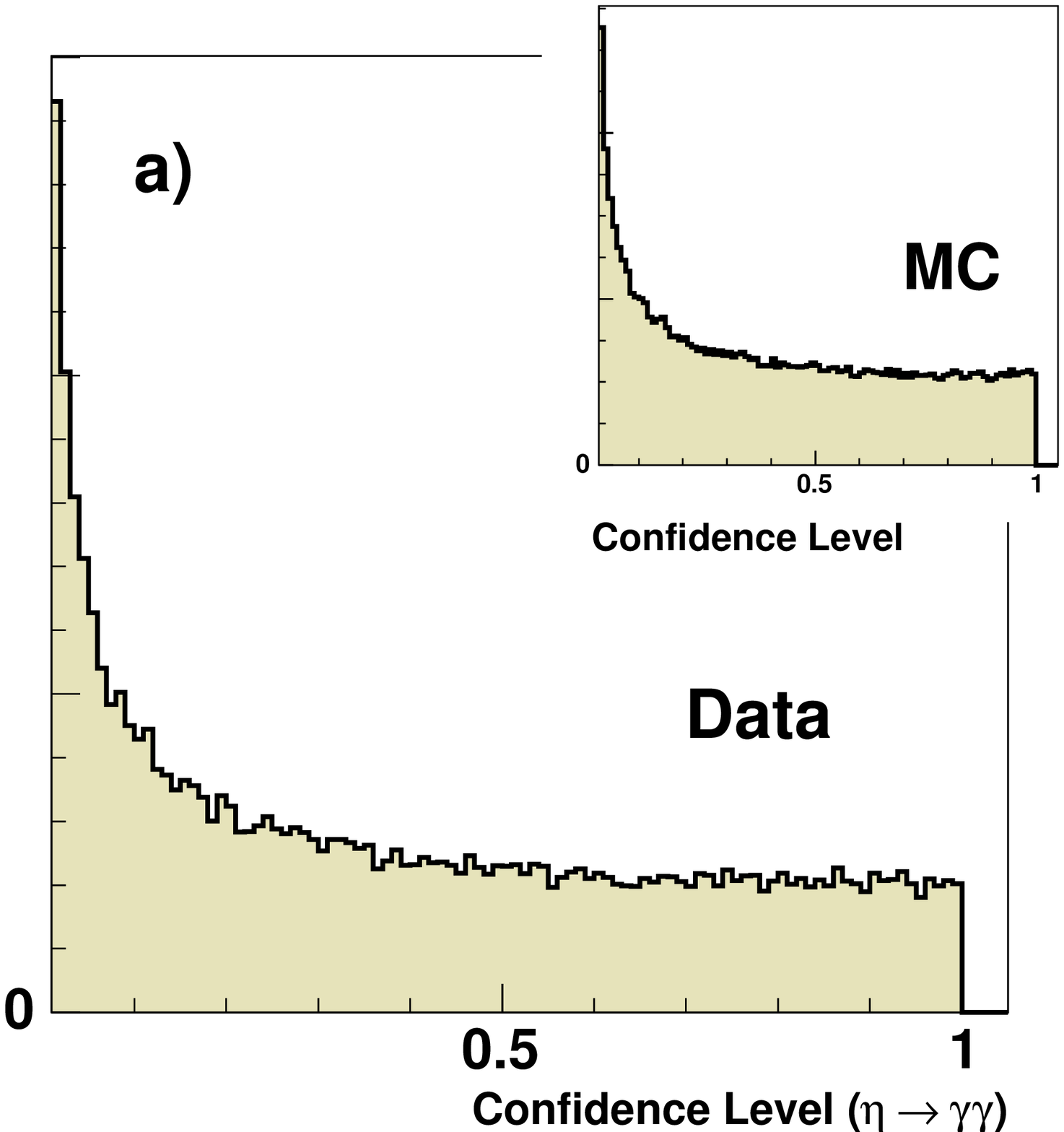,width=.23\textwidth} \hfill
    \epsfig{file=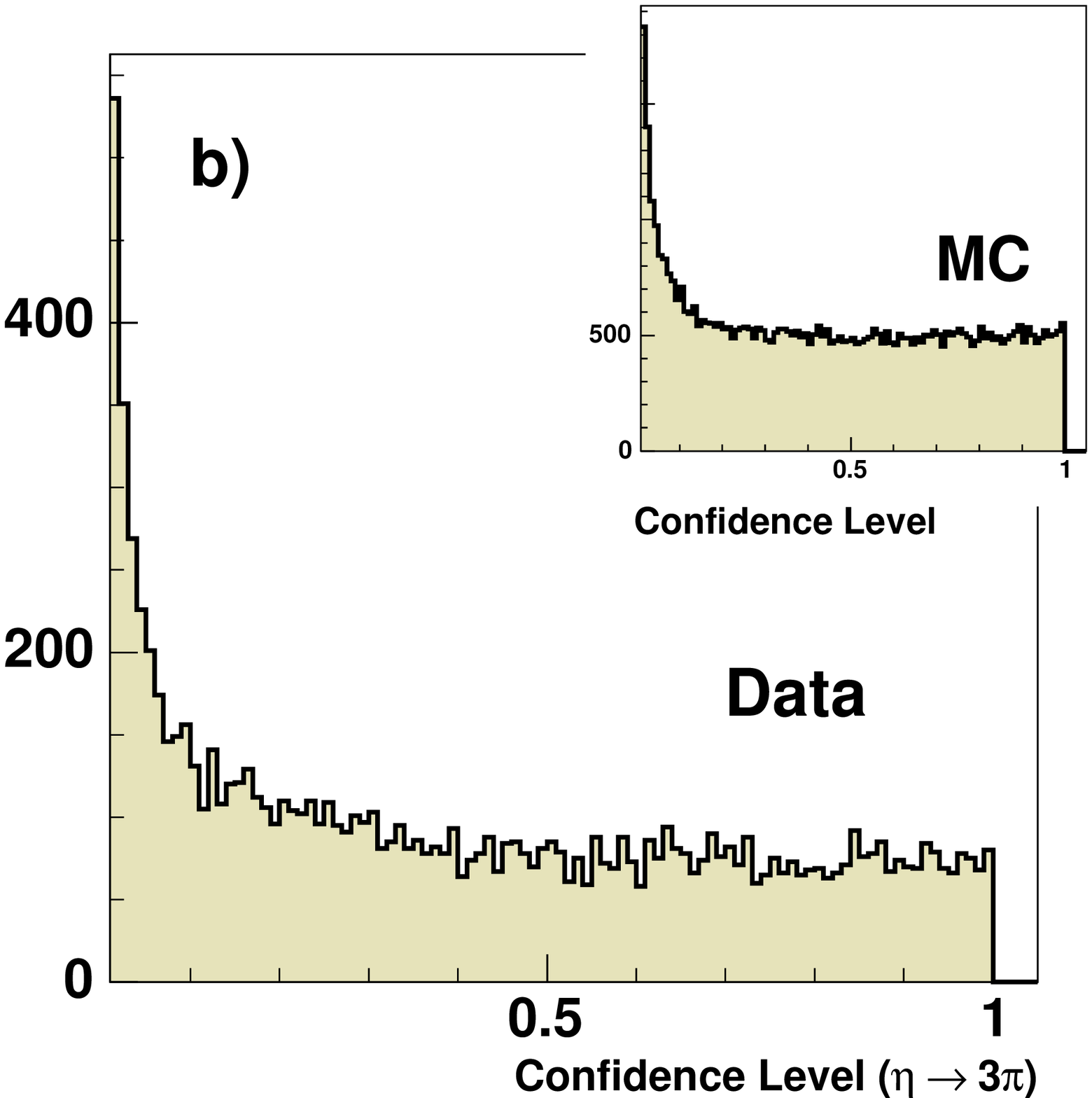,width=.23\textwidth} \hfill
    \epsfig{file=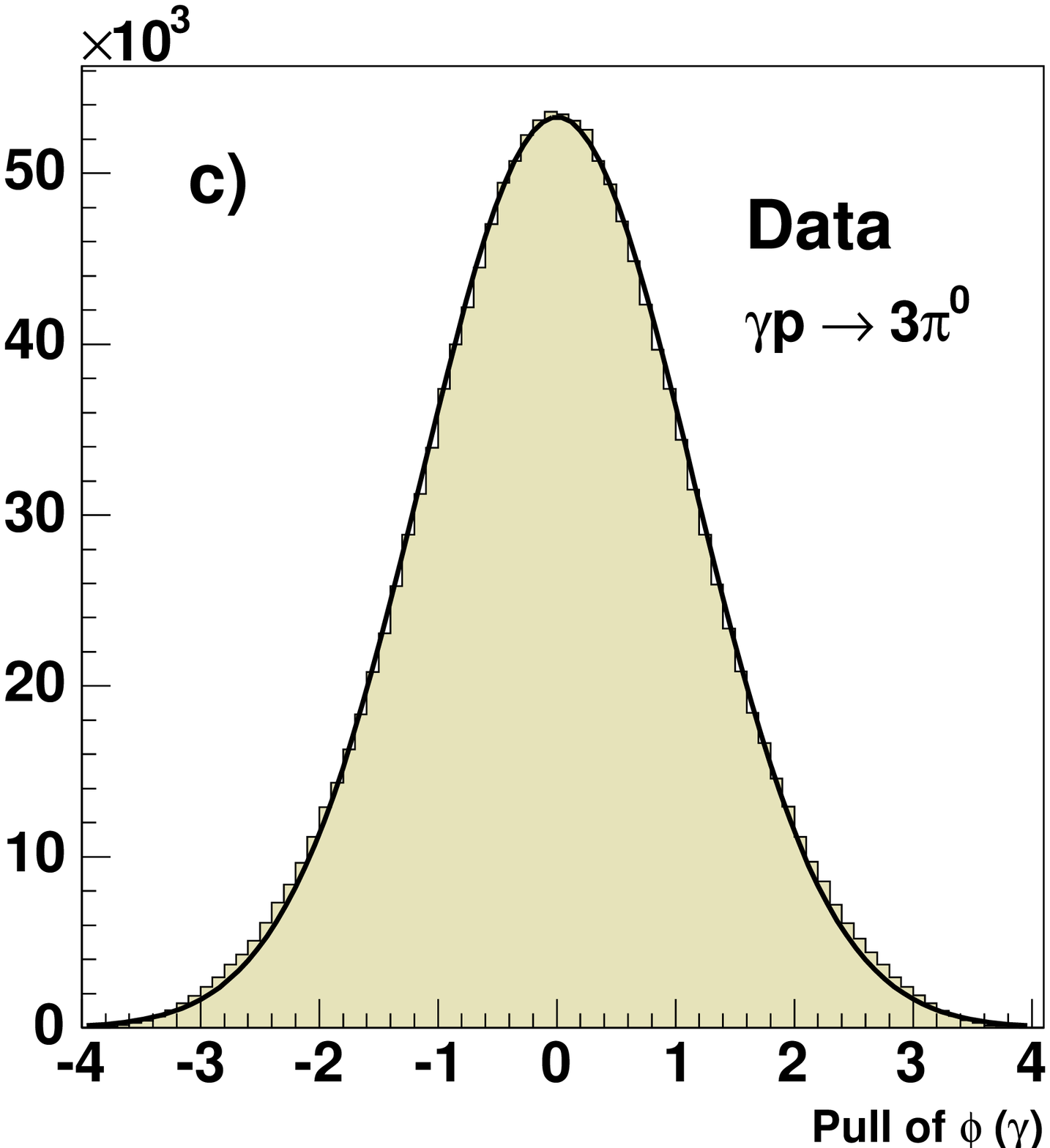,width=.23\textwidth} \hfill
    \epsfig{file=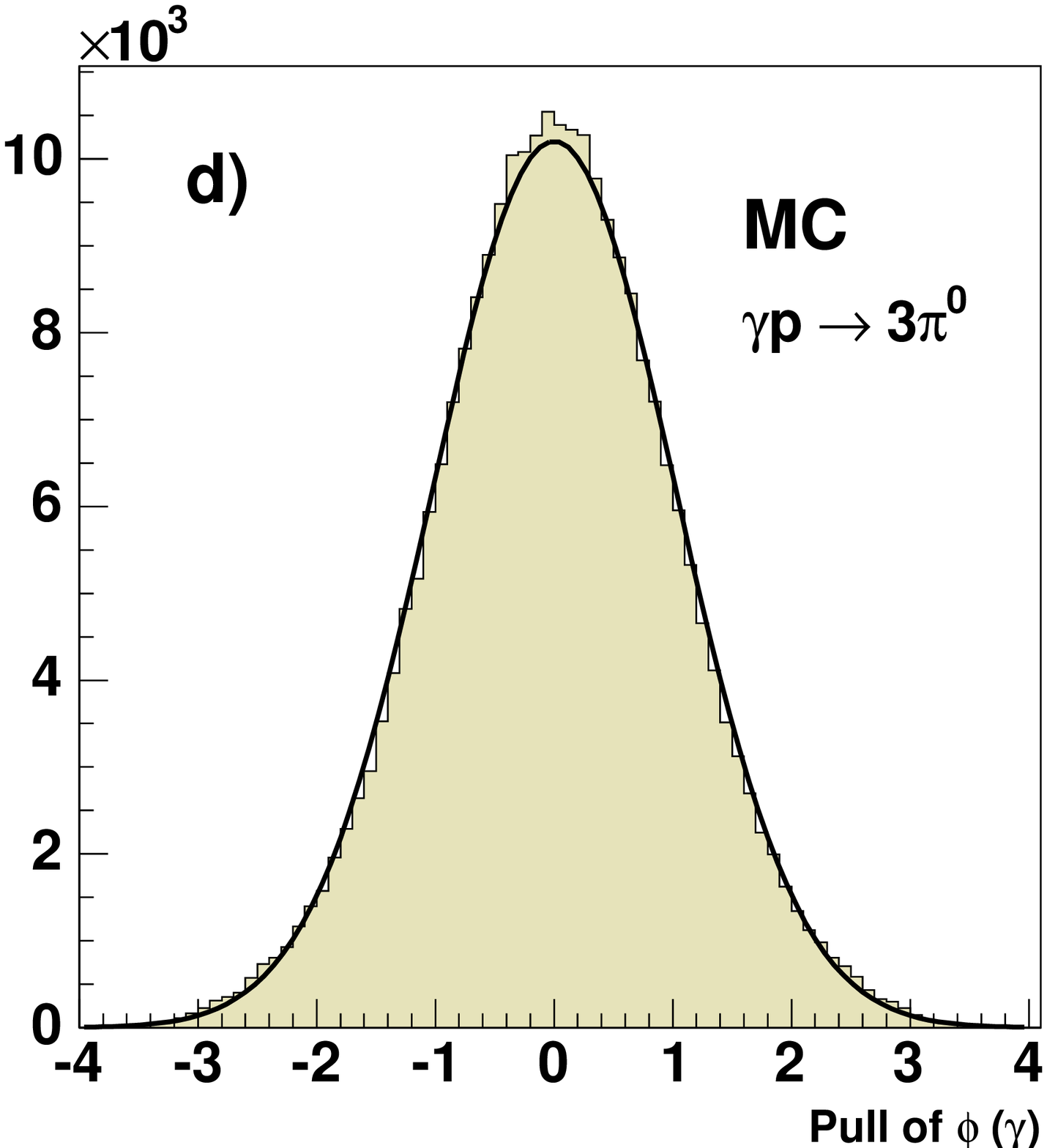,width=.23\textwidth}
  %\end{tabular}
  \caption{\label{cl} Distributions of confidence levels resulting from
    a 1C kinematic fit of a {\it true} data event sample with two photons
    in the final state to the hypothesis $\gamma p\to p\gamma\gamma$ (a) 
    and a 4C fit of a 6-photon event sample to the hypothesis
    $\gamma p\to p 3\pi^0$ (b) for $E_{\gamma}\,\in\,[1600,~1650]$~MeV. 
    The insets show the same for Monte-Carlo data. The strong rise near zero 
    is due to poorly reconstructed signal events. Examples of pull-distributions 
    for the fit parameter~$\phi$ of the final-state photons are shown in (c) 
    for data and in (d) for Monte-Carlo events for the kinematic fit to 
    $\gamma p\to p 3\pi^0$ integrated over the full kinematic range. The 
    mean and $\sigma$~values are 0.0036 and 0.00078 as well as $\sigma = 1.14$ 
    and 1.03, respectively.}
\end{figure*}

The photons hit the liquid hydrogen target in the center of the
Crystal-Barrel (CB) calorimeter. The target cell (5~cm in length,
3~cm in diameter) was surrounded by a scintillating-fiber
detector~\cite{Suft:2005cq}, which provided an unambiguous impact point 
for charged particles (due to the arrangement of its three layers) leaving
the target. The CB-calorimeter in its CBELSA/TAPS configuration of
2002/2003 consisted of 1290 CsI(Tl) crystals with a length of 16~$X_R$. 
The modules have an excellent photon detection efficiency; a detailed 
description can be found in~\cite{Aker:1992}. For this series of experiments, 
the (downstream) rings 11-13 were removed to combine the detector with TAPS 
in the forward direction. The CB-calorimeter covered the complete azimuthal 
angle and polar angles from $30^\circ$ to $168^\circ$. All crystals are of 
trapezoidal shape pointing to the center of the 
target~(Fig.~\ref{Figure:CB-Luzy-H2}, top).

The TAPS detector consisted of 528 hexagonal BaF$_2$ crystals with
a length of about 12~$X_R$. It was configured as an hexagonal wall 
serving as the forward end cap of the Crystal-Barrel calorimeter
(Fig.~\ref{Figure:CB-Luzy-H2}, bottom). TAPS provided a high granularity 
in the forward direction covering polar angles between $5^\circ$ and 
$30^\circ$ (full $\phi$ coverage). A 5~mm thick plastic scintillator 
in front of each TAPS module allowed the identification of charged 
particles. The combination of the Crystal-Barrel and TAPS calorimeters 
covered 99\,\% of the $4\pi$ solid angle and served as an excellent 
setup to detect multi-photon final states.

The fast response of the TAPS modules provided the first-level trigger.
The second-level trigger was based on a cellular logic (FACE), which
determined the number of clusters in the barrel. The trigger required
either two hits above a low-energy threshold in TAPS (LED-low) or one 
hit above a higher-energy threshold in TAPS (LED-high) in combination 
with at least one FACE cluster. The shape of the logical segmentation 
for the TAPS trigger is shown in (Fig.~\ref{Figure:CB-Luzy-H2}, bottom).

The beam-monitor placed at the end of the beam line provided valuable
information on the beam intensity (photons not interacting in the H$_2$
target) used for the determination of the photon flux. This total
absorption $\check{\rm C}$erenkov counter consisted
of an array of 9 lead glass crystals.

\section{\label{Section:DataAnalysis}Data Analysis}
Data presented here were accumulated from October 2002 until
November 2002 in two run periods with ELSA beam energies of 3.175~GeV.
These data were used to extract differential and total cross sections
for a variety of final states \cite{Castelijns:2007qt,Nanova:2008kr}.
The event reconstruction and selection of the two $\eta$ photoproduction
channels (\ref{ReactionEta1}) and (\ref{ReactionEta2}) as well as the
$\eta\,^\prime$ photoproduction channel (\ref{ReactionEtap}) with incident
photon energies up to 2.55~GeV is presented in this section. The total
number of $\sim\,600,000$ $\eta$ events was observed ($\sim\,422,300$
for $\eta\to\gamma\gamma$ and $\sim\,126,300$ $\eta\to 3\pi^0$) covering
invariant masses from 1510 to 2380~MeV/$c^2$. For the $\eta\,^\prime$
channel, $\sim\,5100$ events were observed covering invariant masses
from 1920 to 2380~MeV/$c^2$. The $\eta\,^\prime$~threshold region,
$M\,\in\,[1896,~1920]$~MeV/$c^2$, was analyzed separately using a 
finer energy binning to better study the threshold behavior.

\subsection{Event Reconstruction}
\label{Subsection:EventReconstruction}
Events with at most one proton and with two or six photons were selected, 
respectively. The charged clusters were identified in TAPS by using the plastic 
scintillators mounted in front of each BaF$_2$ crystal. The efficiencies of 
these (photon)-veto detectors were determined and modelled in the Monte-Carlo 
(MC) program. In the Crystal-Barrel reconstruction, a cluster is assigned to 
a charged particle if the trajectory from the target center to the barrel hit 
forms an angle of less than $20^\circ$ with a trajectory from the target center 
to a hit in the scintillating fiber detector. Proton identification is only 
used to remove it from the list of photon candidates. The proton momentum is 
then reconstructed from event kinematics in ``missing-proton'' kinematic 
fitting. Proton clusters are on average much smaller than photon clusters and 
provide worse angular resolution. The proton momentum direction reconstructed 
from kinematic fitting had to be consistent again with a calorimeter hit when 
a charged cluster was identified. Our Monte-Carlo studies of
reaction~(\ref{ReactionEta1}) show that if a proton was observed in the event, 
the overall misidentification probability is less than 3\,\%.

\begin{table}[b]
\begin{center}
\renewcommand{\arraystretch}{1.4}
\begin{tabular}{|c|l|l|c|}
\hline
reaction & decay mode & constraints & fit \\
\hline
$\:\gamma p\to p\eta\:$ & $\:\eta\to 2\gamma$ & $(E,\vec{p}\,)$ 
\:conservation\: & \:1C\:\\
$\:\gamma p\to p\eta\:$ & $\:\eta\to 3\pi^0\to 6\gamma$\: & \:$(E,\vec{p}\,)$ 
conservation\: & \:1C\:\\
& & \quad + $3\times\pi^{0}$ mass & \:4C\:\\
$\:\gamma p\to p\eta\,^\prime\:$ & $\:\eta\,^\prime\to 2\pi^0\eta\to 6\gamma$\: & 
\:$(E,\vec{p}\,)$ conservation\: & \:1C\:\\
& & \quad + $\pi^{0}\,\eta\,\gamma\gamma$ masses & \:3C\:\\
\hline
\end{tabular}
\caption{Kinematic fits and constraints used in the analysis. The
proton is treated as missing particle. Its momentum is determined from
the kinematic fit. For the $\eta\,^\prime$ reconstruction, only one 
pion mass was imposed in the kinematic fit (3C fit).} 
\label{TableKinematicFits}
\renewcommand{\arraystretch}{1.0}
\end{center}
\end{table}

A kinematic fitter slightly adjusts the measured values within the 
estimated errors by a minimization procedure until they fulfill exactly 
certain constraints expressed in the form of equations, which are based 
on physical conditions like energy and momentum conservation or invariant 
masses. The $\chi^2$ probability or confidence level (CL), which is derived 
from the $\chi^2$ value of the fit, can be used to make judgements and decisions 
about the {\it goodness\/} of the fit and provides an ideal method to judge 
possible final-state hypotheses for an event. For Gaussian-distributed errors
of the measured particle properties, confidence levels should be flat.
{\it Pulls\/} are defined to test the correct determination of the covariance
matrix and are a measure of the displacement of the reconstructed values to 
the fitted values. They are constructed such that a valid distribution of pulls 
will form a normal distribution with a width of one and a mean of zero. Pulls 
are very sensitive to the goodness of the fit. If the width deviates from one, 
the resolution derived in the reconstruction does not reflect the true errors 
and it is necessary to globally scale the measured initial errors in order to 
force the pull distributions to have a width of one. No scaling factors are 
needed in this analysis for events with two photons in the final state. For 
events with six photons in the final state, scaling factors have been determined 
carefully for data and Monte-Carlo events. Typical confidence-level and pull 
distributions are shown in Fig.~\ref{cl}. The CL values were found to be 
sufficiently flat in all photon-energy bins.

\begin{figure}[b!]
  \vspace{-4mm}
  \epsfig{file=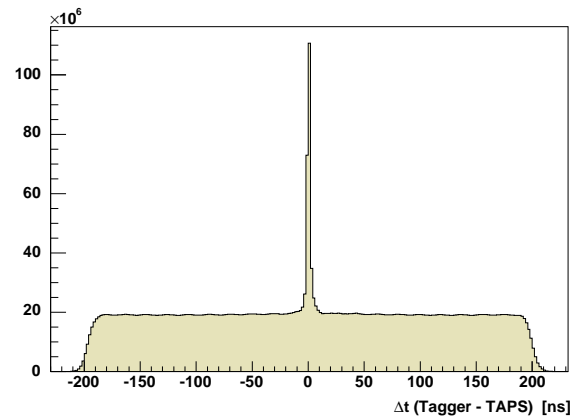,width=0.48\textwidth}
  \caption{\label{FigureTiming} Time difference between photons in TAPS
    (mean value) and electrons in the tagger integrated over all events. A
    prompt coincidence is defined by $-3 < \Delta t < 3$ ns.}
\end{figure}

\begin{figure*}
  \epsfig{file=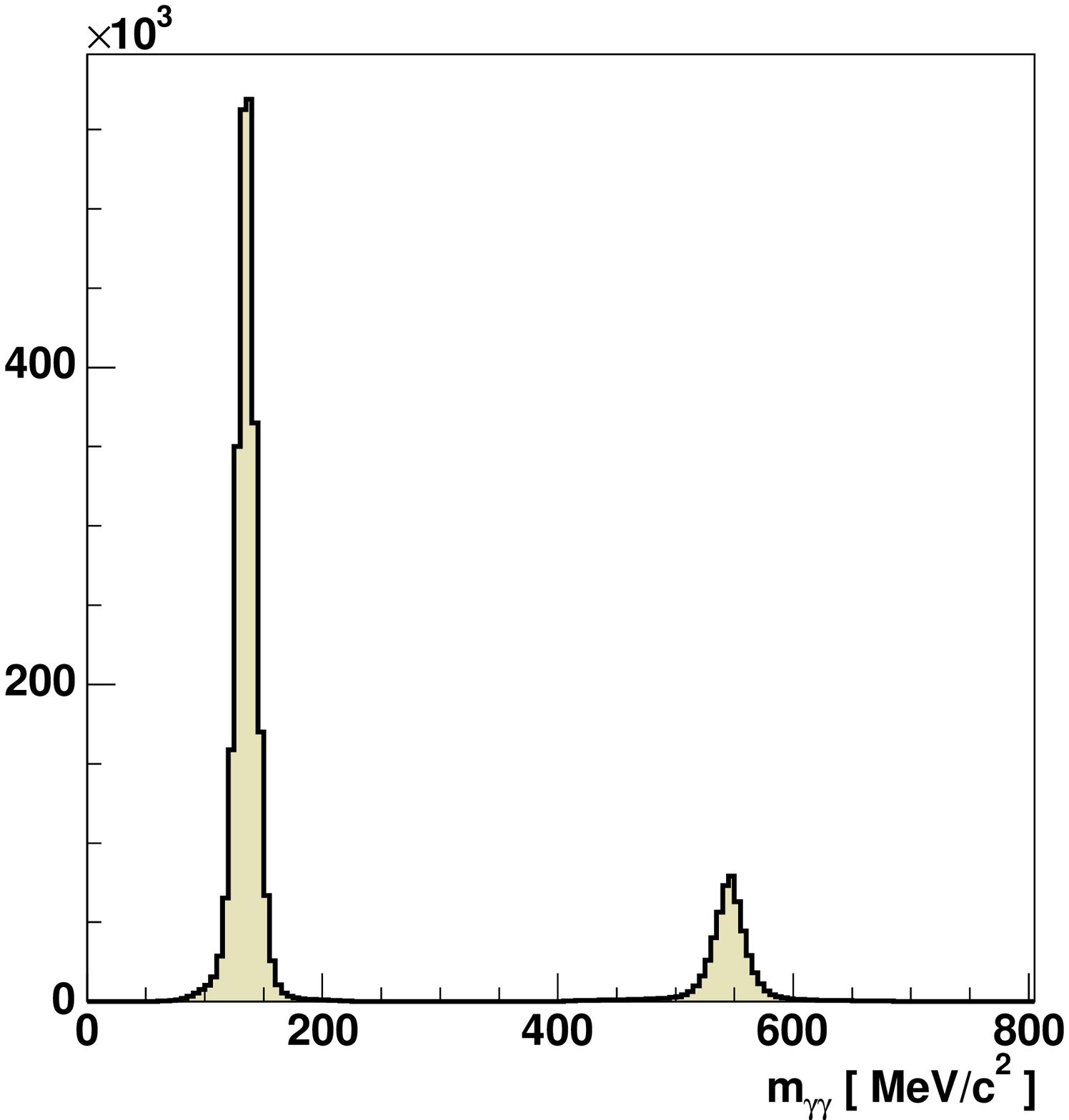,width=.32\textwidth} \hfill
  \epsfig{file=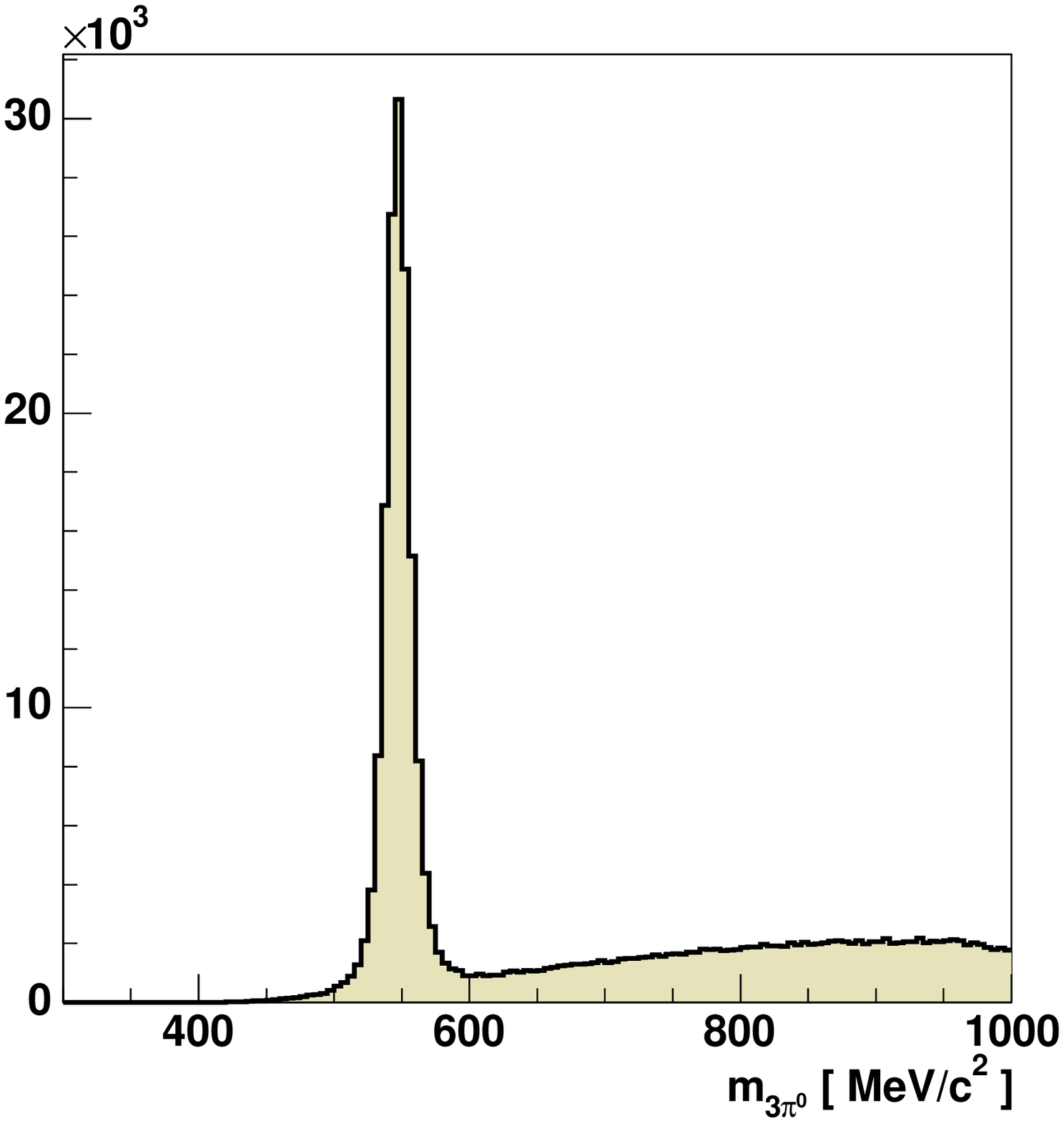,width=.32\textwidth} \hfill
  \epsfig{file=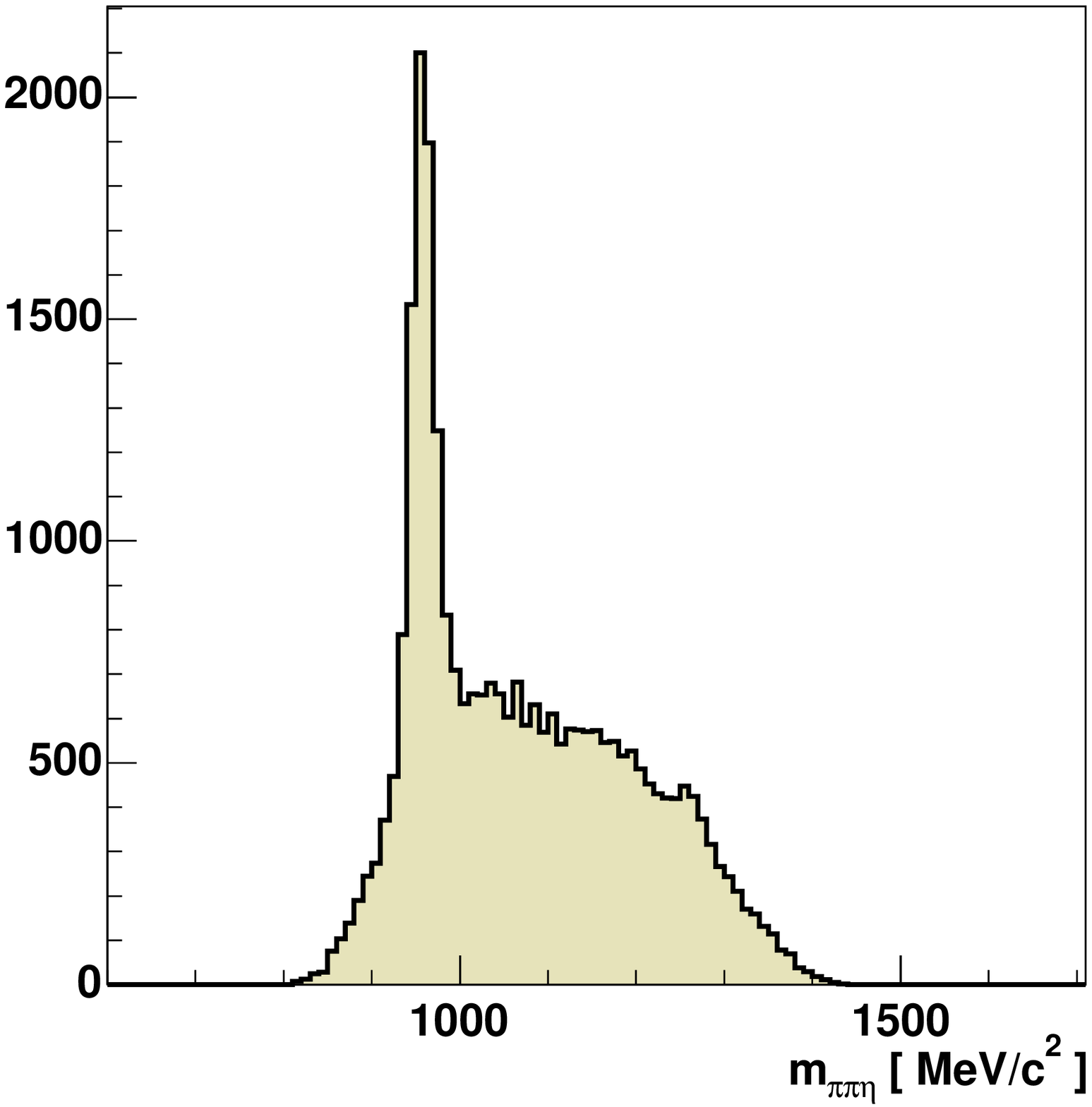,width=.32\textwidth}
  \vspace{-1mm}
  \caption{\label{ggMass} Invariant mass spectra (integrated over all
    photon energies) for the reactions $\gamma p\to p\gamma\gamma$ (left)
    and $\gamma p\to p\pi^0\pi^0\pi^0$ (center); confidence-level cuts 
    were applied at $10^{-2}$ and at $10^{-3}$, respectively. In the 
    2-photon decay mode, the $\pi^0$ and $\eta$ mesons are observed with 
    very little background. The invariant $\pi^0\pi^0\eta$ mass spectrum 
    (right) shows a clear $\eta\,^\prime$ signal and an enhancement 
    at 1250~MeV/$c^2$.}
\end{figure*}

In a first step of kinematic fitting, a consistency check was carried
out by imposing energy and momentum conservation on all events. The
hypothesis
\begin{equation}
  \label{EquationNGammaHypothesis}
  \gamma p\to p\, n_{\gamma}\gamma
\end{equation}

\noindent
was tested, where $n_{\gamma}$ is the number of photons in the final
state, i.e. two for reaction (\ref{ReactionEta1}) and six for
(\ref{ReactionEta2}) and (\ref{ReactionEtap}). Energy and momentum
conservation provides four equations which any event due to reaction
(\ref{EquationNGammaHypothesis}) must satisfy. Thus, the proton
three-momentum can be left ``missing'' and reconstructed from other
observables, still retaining one constraint (1C) provided by the photon
energies and directions. Table~\ref{TableKinematicFits} summarizes the 
hypotheses used to select events for $\gamma p\to p\eta$ and
$\gamma p\to p\eta\,^\prime$.

\begin{figure}[b]
  \vspace{-4mm}
  \epsfig{file=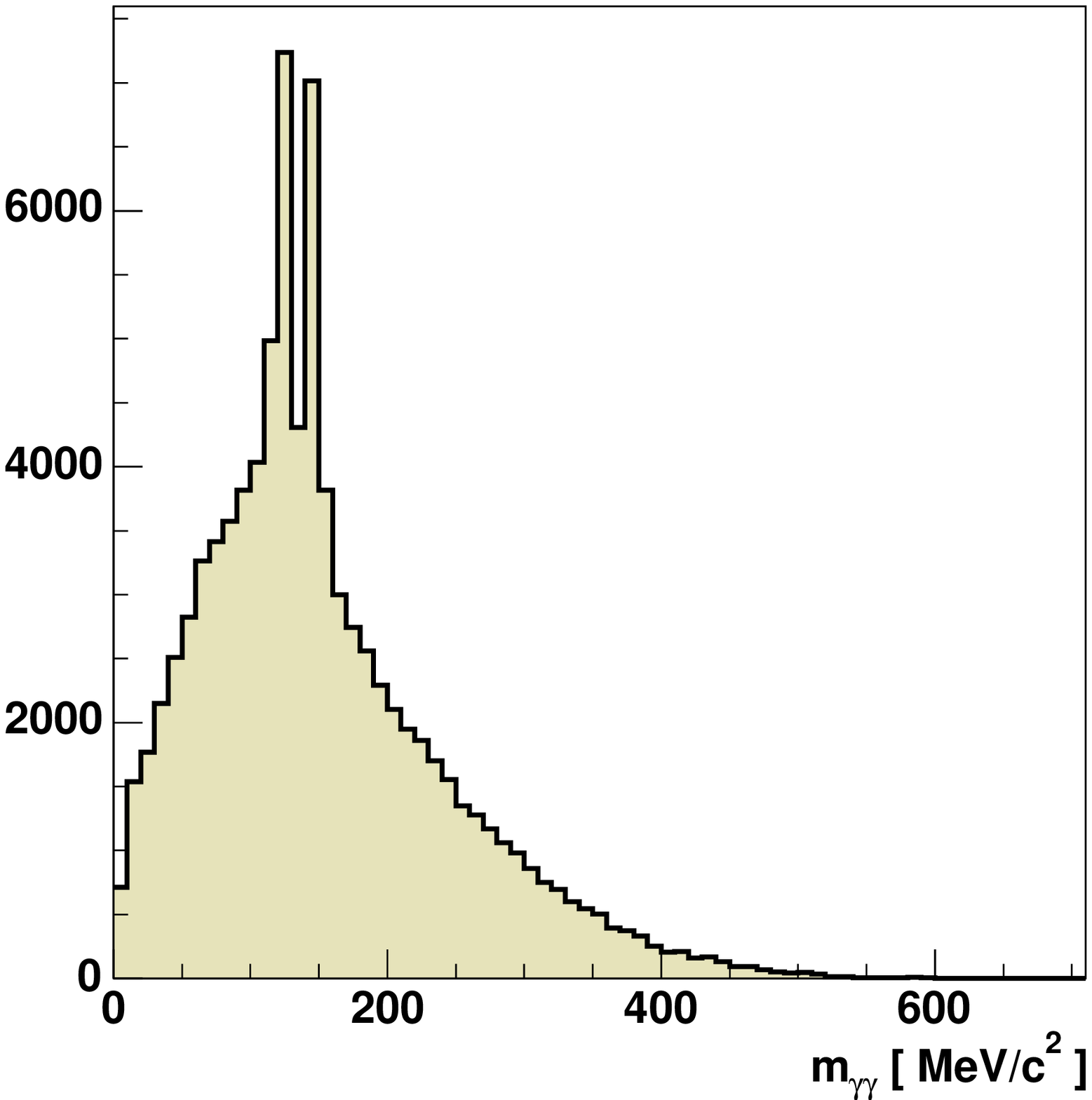,width=0.23\textwidth}
  \epsfig{file=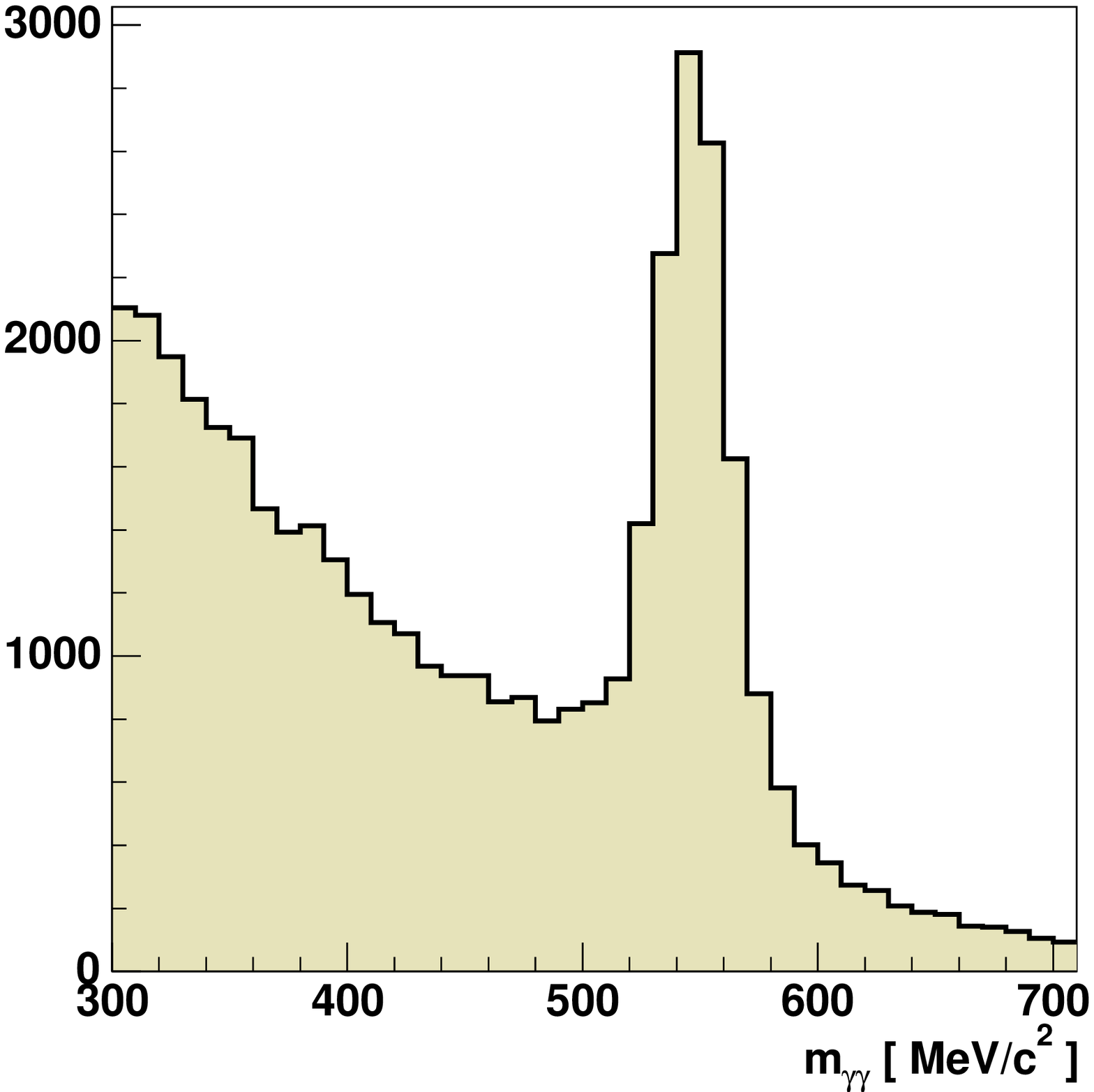,width=0.23\textwidth}
  \caption{\label{Figure:pi0pi0gg} Invariant $\gamma\gamma$ mass spectra 
    (integrated over all incoming photon energies) for the reactions 
    $\gamma p\to p\pi^0\eta\gamma\gamma$ (left) and $\gamma p\to p\pi^0
    \pi^0\gamma\gamma$ (right). The $\pi^0$ signal shows a double-peak
    structure since the ``better'' $\pi^0$ was found in the kinematic
    fit.}
\end{figure}

A prompt coincidence between a photon in TAPS and an electron in the 
tagger was required to reduce time accidental background. Random time 
coincidences underneath the prompt peak (Fig.~\ref{FigureTiming}) were 
subtracted by performing the exact same selection procedure for events 
outside the prompt time coincidence window. Fig.~\ref{ggMass} (left) 
shows the invariant $\gamma\gamma$ mass (time-accidental background
subtracted) for kinematically-fitted two-photon events. A confidence level
cut at $10^{-2}$ was applied. Clear peaks for the $\pi^0$ and $\eta$
mesons are visible. The background underneath the peaks depends on
kinematics and is on the average smaller than 4\,\%, but can be up to 
15\,\% at high energies and in the most forward angle bins. In addition 
to energy and momentum conservation, three $\pi^0$~mass constraints were 
imposed on events due to $\gamma p\to p\eta\to p\,3\pi^0\to p\,6\gamma$ 
(4C~kinematic fit) reducing significantly combinatorial background. A 
confidence level cut at $10^{-3}$ was applied. The invariant $3\pi^0$ mass 
is shown in Fig.~\ref{ggMass} (middle). The $\eta$ peak is visible above 
a small combinatorial background. The uncertainty in signal lost due to 
CL-cuts in data compared to Monte Carlo is estimated to be less than~3\,\%.

In a final step of the $\eta$ analysis, mass cuts were applied in the
two-photon spectrum and in the $3\pi^0$ spectrum. The width of the
$\eta$~peak varied as a function of incident photon energy between
$\sim 12$~MeV/$c^2$ at the lowest energies and $\sim 17$~MeV/$c^2$ at
the highest energies. The remaining background underneath the
$\eta\,_{\gamma\gamma}$ and $\eta\,_{3\pi^0}$ peaks was subtracted 
using side bins for every ($E_\gamma,\,{\rm cos}\,\theta$) bin.

For the selection of $\eta\,^\prime$ events, the hypothesis $\gamma p\to 
p\pi^0\eta\gamma\gamma$ was tested in addition to reaction
(\ref{EquationNGammaHypothesis}) with a confidence-level cut at $10^{-2}$.
The remaining invariant $\gamma\gamma$ mass is shown in 
Fig.~\ref{Figure:pi0pi0gg}~(left). The second $\pi^0$ was reconstructed with 
$110 < m_{\pi^0\to\gamma\gamma} < 160$~MeV/$c^2$. Fig.~\ref{ggMass} (right) shows 
a clear peak for the $\eta\,^\prime$ in the invariant $2\pi^0\eta$ mass spectrum. 
Moreover, an interesting enhancement is visible at 1250~MeV/$c^2$ giving
rise to a possible contribution of the controversial meson $\eta(1295)$  
and/or the $f_1(1285)$. Further studies of this signal are statistically 
challenging and are not discussed further here. Events due to 
reaction~(\ref{ReactionEtap}) were finally selected with 
$910 < m_{2\pi^0\eta} < 1010$~MeV/$c^2$. The remaining background was 
determined in fits to the $\eta\,^\prime$ peaks. An alternative way of 
reconstructing $\eta\,^\prime$ events via $\gamma p\to p\pi^0\pi^0\gamma\gamma$ 
was used for systematic checks. The invariant $\gamma\gamma$ mass is shown in 
Fig.~\ref{Figure:pi0pi0gg}~(right).

\subsection{\label{Section:MonteCarloSimulations}Monte-Carlo simulations}
The performance of the detector was simulated in GEANT3-based Monte-Carlo 
studies. The program package used for CBELSA/TAPS is built upon a program 
developed for the CB-ELSA experiment. The Monte-Carlo program reproduces 
accurately the response of the TAPS and Crystal-Barrel crystals when hit 
by a photon. For charged particles, the detector response is known to a 
lower precision but still reasonably well understood.

The acceptance for the reactions (\ref{ReactionEta1})-(\ref{ReactionEtap})
was determined by simulating events, which were evenly distributed over the 
available phase space. The Monte-Carlo events were analyzed using the same 
reconstruction criteria, which were also applied to the (real) measured data.
The same hypotheses were tested in the kinematic fits and events selected with 
the same confidence level cuts. The acceptance is defined as the ratio of the 
number of generated to reconstructed Monte-Carlo events
\begin{equation}
  A_{\gamma p\to p\,X}=\frac{N_{\rm rec,MC}}{N_{\rm gen,MC}}
     \qquad (X=\eta,\,\eta\,^\prime).
\end{equation}

The trigger required either two hits above a low-energy threshold in TAPS
(LED low -- Leading-Edge Discriminator threshold) or one hit above a
higher-energy threshold in TAPS (LED high) in combination with at least
one cluster in the Crystal-Barrel calorimeter. The second-level trigger
used a fast cluster encoder (FACE) based on a cellular logic to define
the number of contiguous Crystal-Barrel clusters. The decision time depended on the
complexity of the hit distribution in the Crystal-Barrel and was typically
4 $\mu$s. In case of an event rejection, a fast reset was generated, which
cleared the readout electronics in 5 $\mu$s. Otherwise the readout of the
full event was initiated with typical readout times of 5-10 ms. To
properly simulate the detector response, FACE and TAPS-LED thresholds
had to be determined from the data for all crystals. Given the different
response characteristics of protons and photons in BaF$_2$ crystals,
protons experience slightly different LED-thresholds than photons. For
this reason, we have corrected the measured proton energy according to
$0.8\cdot E_{\rm p} + 30$~MeV, which is derived from available proton
times in TAPS and from Monte-Carlo studies. At the reaction threshold,
when the proton is required in the (TAPS) trigger, corrections are small.
Above about 1~GeV in the incoming photon energy, the proton trigger is
not relevant. Our understanding of the threshold function is fair to good 
and reasonably well reproduced in the trigger simulation.

\section{Determination of Cross Sections}
\label{Section:DeterminationCS}

The differential cross sections for this analysis are determined
according to
\begin{align}
\frac{\rm d\sigma}{\rm d\Omega}&=\frac{N_{X\rightarrow
n_{\gamma}\gamma}}{A_{X\rightarrow
n_{\gamma}\gamma}}\medspace\frac{1}{N_{\gamma}\rho_{\rm
t}}\medspace\frac{1}{\Delta\Omega}\medspace\frac{1}
{\frac{\Gamma_{X\rightarrow n_{\gamma}\gamma}}{\Gamma_{\rm
total}}}, \label{gldsigdom}
\vspace{-10mm}
\end{align}
where

\begin{table}[H]
\begin{tabular}{rl}
$\rho_{\rm t}$\,: & target area density\\
N$_{X\to n_\gamma \gamma}$\,: & number of reconstructed data events\\
 & in an ($E_\gamma$,\,cos\,$\theta_{cm}$) bin\\
$N_\gamma$\,:& number of photons in an ($E_\gamma$) bin\\
A$_{X\to n_\gamma \gamma}$\,: & acceptance in an ($E_\gamma$,\,cos\,$\theta_{cm}$) bin\\
$\Delta\Omega$\,: & solid-angle interval $\Delta\Omega= 2\pi\Delta cos\,(\theta_{cm})$\\
$\frac{\Gamma_{X\to n_\gamma \gamma}}{\Gamma_{total}}$\,: & decay branching fraction.
\end{tabular}
\end{table}

The target area density, i.e. the number of atoms in the target
material per cross-sectional area (orthogonal to the photon beam),
is given by
\begin{equation}
\rho_{\rm t} = 2\,\frac{\rho({\rm H}_2) N_A L}{M_{mol}({\rm H}_2)} = 2.231\cdot 10^{-7}
\mu {\rm b}^{-1}\,,
\end{equation}
where $\rho({\rm H}_2) = 0.0708$ g/cm$^3$ is the density and $M_{mol} =
2.01588$ g/mol the molar mass of liquid H$_2$. $N_A = 6.022\cdot 10^{23}$
mol$^{-1}$ is the Avogadro number and $L = 5.275$ cm the length of the
target cell. The factor of two accounts for the molecular composition
of hydrogen (H$_2$).

The cross sections were extracted independently for both $\eta$~decay modes, 
$\eta\to\gamma\gamma$ as well as $\eta\to 3\pi^0$, and then averaged (weighted 
with their errors) based on the observed good agreement (Fig.~\ref{ratioDecayModes}).
The number of events in an ($E_\gamma$,\,cos\,$\theta_{cm}$) bin comprises events 
with two or three final-state particles (at least $2\gamma$'s) or six or seven 
``particles'' (at least $6\gamma$'s), respectively. The proton can be ``missing'', 
but events with and without detected proton are treated in the same way in the 
event reconstruction. At threshold, the event kinematics requires that the proton
is used in the (TAPS) trigger. Thus, the threshold function for the detection of 
low-energy protons needs to be reasonably well understood.

The interval of the solid angle is given by $\Delta\Omega=2\pi\Delta\cos(\theta_{\rm cm})$ 
with $\Delta \cos(\theta_{\rm cm})$ describing the width of the angular bins. It 
was chosen to be $0.1$ for the $\eta$ data presented here. Energy bins were defined 
by considering statistics and ensuring a good comparability with other experiments. 
A total of 34~bins is presented in energy steps of 50~MeV for $E_{\gamma}\,\in\,
[850,~2550]$~MeV. For the $\eta\,^\prime$ data, $\Delta \cos(\theta_{\rm cm})$ was 
chosen to be 0.2. A total of 20 bins is presented in energy steps of 50~MeV for 
$E_{\gamma}\,\in\, [1500,~2550]$~MeV.

The number of observed $\eta$ and $\eta\,^\prime$ mesons needs to be corrected for 
unseen decay modes. Partial-decay branching fractions used to correct the measured 
cross-sections are taken from~\cite{Amsler:2008zz}: BR$(\eta\to 2\gamma) = 0.3931\pm 
0.002$, BR$(\eta\to 3\pi^0\to 6\gamma) = 0.3256\pm 0.0023$, and BR$(\eta\,^\prime\to 
2\pi^0\eta\to 6\gamma) = 0.207\pm 0.012$.

\subsection{Normalization\label{Subsection:Normalization}}
The tagging hodoscope consisted of 480 scintillating fibers above 14
partially overlapping scintillation counters (tagger bars). The photon flux
was measured directly in the experiment and determined according to
\begin{equation}
N_\gamma\,=\, N\,^{\rm fiber}_{\rm scaler} \cdot \alpha \cdot P_\gamma\,,
\end{equation}

\noindent
where $N\,^{\rm fiber}_{\rm scaler}$ are the {\it free} hardware counts
for the individual fibers corrected for the life-time of the detector.
The parameter $\alpha$ accounts for the (fiber)-cluster reconstruction
in the tagger, which has to be performed in the same way as for real
hadronic events. The photon definition probability or $P_\gamma$ denotes
the probability that a real photon is emitted along the beam axis in the
tagger and traverses the liquid hydrogen target. The scalers are recorded
in {\it scaler} events, which were accumulated with a {\it minimum-bias}
trigger at a rate of 1\,Hz during the regular data taking. This trigger
required only a hit in the tagger and was thus independent of hadronic
cross sections. The parameter $P_\gamma$ is determined from Tagger-Or-Runs
$-$ separate data runs utilizing a {\it minimum-bias} trigger.

The total error of the photon flux is assumed to be dominated by the
$P_\gamma$ error and depends strongly on the efficiency of the beam monitor
at the end of the beam line (Fig.~\ref{Figure:Experiment}). $P_\gamma$ was
determined to $0.639\pm 0.002\,_{\rm stat.}\pm 0.05\,_{\rm sys.}$ by varying 
the background subtraction of non-coincident tagger-beam monitor hits. This 
value is consistent with determinations from multiple Tagger-Or-Runs at 
different incoming photon rates. An overall error of 10\,\% has been assigned
to the photon flux determination.

\subsection{Systematic Uncertainties\label{Subsection:SystematicUncertainties}}

The statistical errors are determined from the number of events in each 
($E_\gamma$,\,cos\,$\theta_{cm}$) bin. Statistical errors are shown for all 
data points; systematic uncertainties are given as error bands at the bottom 
of each plot.

\begin{figure}
  \vspace{-1mm}
  \epsfig{file=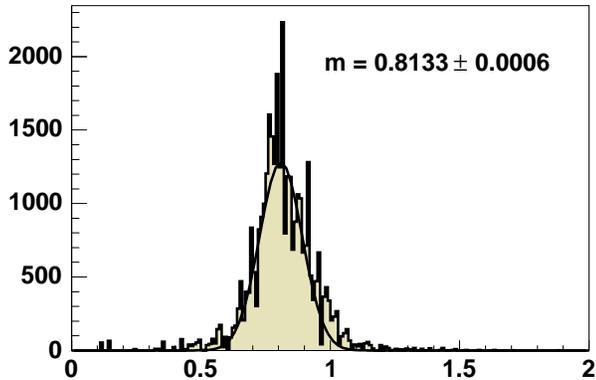,width=0.48\textwidth}
  \vspace{-3mm}
  \caption{\label{ratioDecayModes} Ratio $\Gamma_{\eta\to 3\pi^0}$\,/
    \,$\Gamma_{\eta\to 2\gamma}$ for each ($E_\gamma$,\,cos\,$\theta_{cm}$)
    bin. The values are weighted with their squared inverse errors.}
  \vspace{-3mm}
\end{figure}

\begin{figure}
  \vspace{-5.7mm}
  \epsfig{file=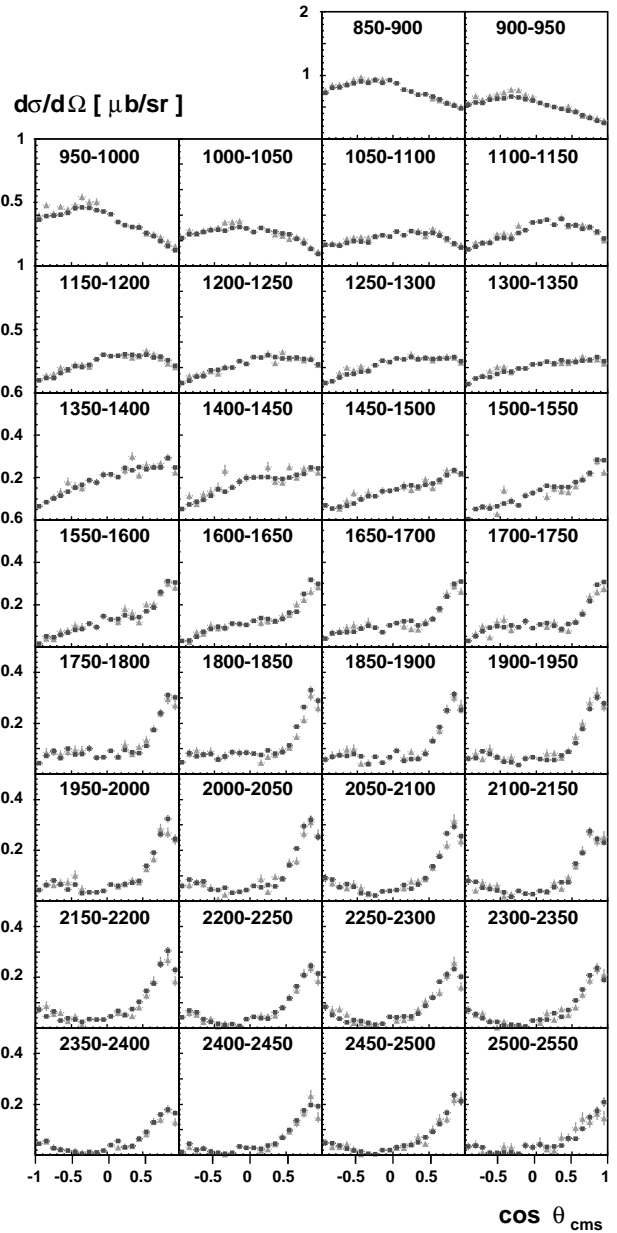,width=0.50\textwidth}
  \vspace{-12mm}
  \caption{\label{Figure:EtaDCS_1} Differential cross sections for $\gamma p\to
      p\eta$ using the $\eta\to 2\gamma$ ({\tiny\color{gray07}$\blacksquare$})
      and $\eta\to 3\pi^0\to 6\gamma$  ({\footnotesize\color{gray04}$\blacktriangle$})
      decay channels. Only statistical errors are assigned to the data points. In some
      bins, the acceptance for $\eta\to 3\pi^0$ is small $(<\,5\,\%)$ and thus,
      corresponding data points are not shown. The figure shows the excellent agreement
      of both reactions.}
\end{figure}

Sources of the systematic errors are uncertainties in the exact position of 
the liquid hydrogen target and a possible offset of the photon beam. The 
position of the target cell was determined from kinematic fitting by comparing 
the off-zero displacement of different pull distributions to Monte-Carlo 
simulations. It was found to be shifted upstream by 0.65~cm~\cite{van Pee:2007tw}. 
The corresponding systematic errors were determined by varying the target position 
in the Monte Carlo $(\pm 1.5~{\rm mm})$ and evaluating changes in the re-extracted 
differential cross sections. The errors show an angular dependence, but are 2-3\,\% 
on the average and $\leq 5$\,\% at most around cos\,$\theta_{cm} = 0$. The photon 
beam was assumed to be shifted by less than 2~mm off axis at the target position. 
The errors of the decay branching fractions are negligible. The uncertainty of the 
proton trigger has been determined from the small disagreement of the differential 
$\eta$ cross sections using the $\eta\to 2\gamma$ and $\eta\to 3\pi^0\to 6\gamma$ 
decay channels for $E_\gamma < 1$~GeV and cos~$\theta_{cm} < 0.0$~(Fig.~\ref{Figure:EtaDCS_1}).

\begin{figure*}
  \vspace{-2mm}
  \begin{center}
    \epsfig{file=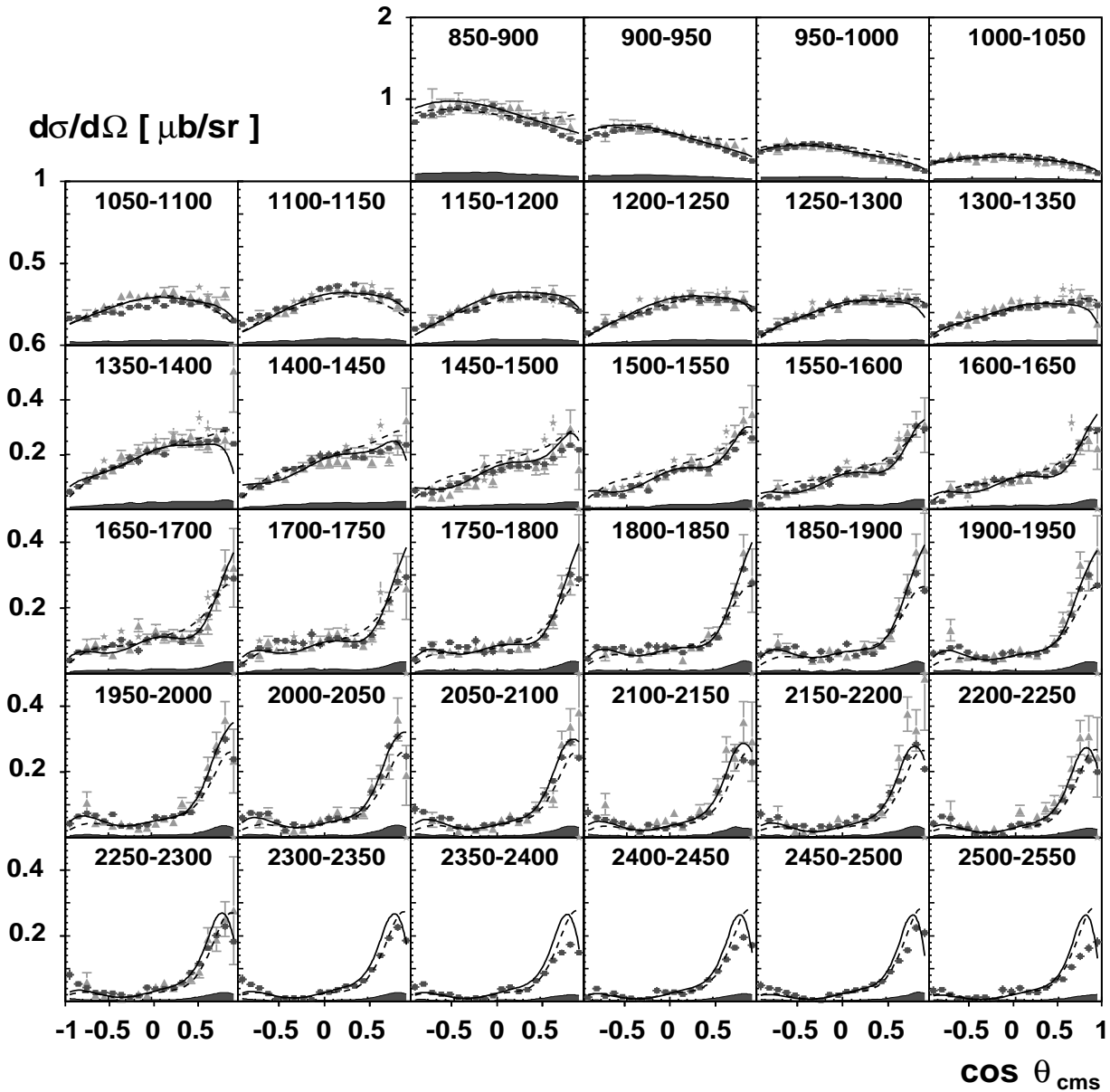,width=1.0\textwidth,height=0.95\textwidth}
  \end{center}
  \vspace{-6mm}
  \caption{\label{Figure:EtaDCS_2} Differential cross sections for
    $\gamma p\to p\eta$ ({\tiny\color{gray07}$\blacksquare$})
    using the combined data set of $\eta\to 2\gamma$ and $\eta\to 3\pi^0\to
    6\gamma$ events. For comparison, CB-ELSA data~\cite{Crede:2003ax} are
    represented by ({\footnotesize\color{gray04}$\blacktriangle$}) and CLAS
    data by ({\footnotesize\color{gray04}$\bigstar$}). The solid line shows
    our previous PWA solution~\cite{Crede:2003ax} and the dashed line represents 
    the SAID model~\cite{Arndt:2002xv}. The data points include statistical errors
    only; the total systematic error is given as error bands at the bottom of each plot.}
\end{figure*}

The reconstruction of neutral mesons and the identification of final states
requires a sequence of cuts including the use of kinematic fitting. As discussed 
in the following section, the reconstruction of $\eta$ mesons from final states 
with two and six photons leads to compatible results. This fact emphasizes a good 
understanding of the detector response to multi-photon final states. An overall
$\pm 5.7\,\%$ error is assigned to the reconstruction efficiency as determined 
in~\cite{Amsler:1993kg}. An additional 3\,\% systematic error accounts for the
slightly different effects of confidence-level cuts on data and Monte-Carlo
events. All these errors are added quadratically to give the 
total systematic error. Moreover, the $\eta\,^\prime$ systematic error receives 
an additional contribution from an alternative way of reconstructing events via 
$\gamma p\to p\pi^0\pi^0\gamma\gamma$~(Sec.~\ref{Subsection:EventReconstruction}).

\section{Experimental Results \label{Section:Results}}

\subsection{Differential Cross Sections $\boldsymbol{d\sigma/d\Omega}$
for $\boldsymbol{\gamma p\to p\eta}$ at an Electron-Beam Energy of
$\boldsymbol{E_{{\rm e}^-} = 3.18}$~GeV}
\label{Subsection:CrossSectionsEta}

\begin{figure*}[t!]
  \epsfig{file=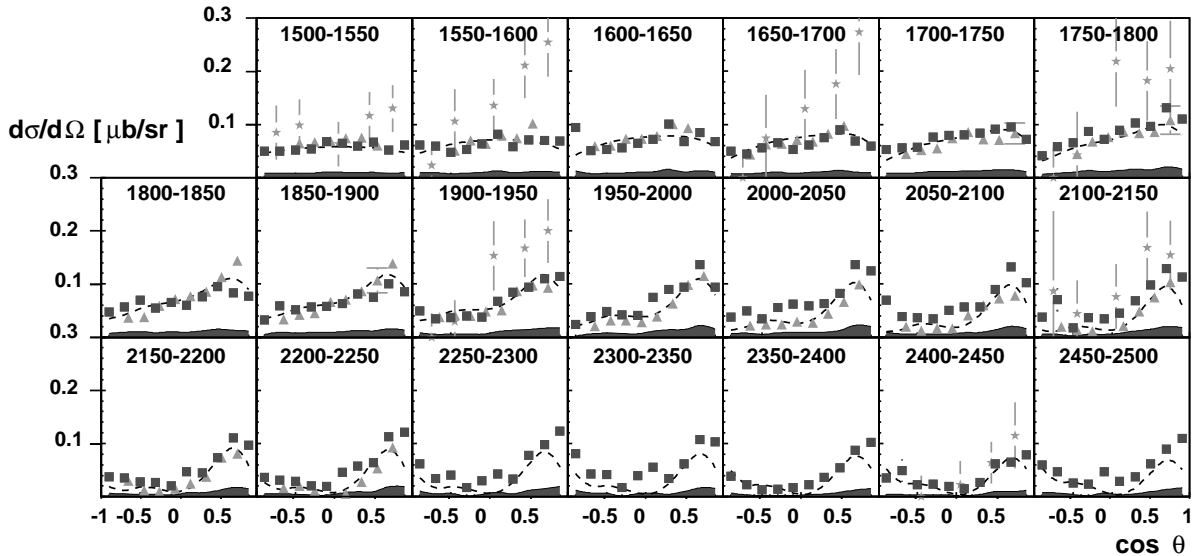,width=0.95\textwidth}
  \caption{\label{Figure:EtapDCS} Differential cross sections for the reaction 
    $\gamma p\to p\eta\,^\prime\to 2\pi^0\eta\to p6\gamma$ ({\tiny\color{gray07}
    $\blacksquare$}) using 50-MeV~wide energy bins and cos\,$\theta_{\rm c.m.}^{\,
    \eta\,^\prime}$ bins of width~0.2. The data cover the full angular range; 
    energies are given in MeV. For comparison, data are shown from 
    SAPHIR~\cite{Plotzke:1998ua}~({\footnotesize\color{gray04}$\bigstar$}) and 
    CLAS~\cite{Dugger:2005my}~({\footnotesize\color{gray04}$\blacktriangle$}). 
    The SAPHIR data are based on only 250~events and thus, have large error
    bars. The dashed line represents the SAID~model~\cite{Arndt:2002xv}.}
\end{figure*}

Fig.~\ref{Figure:EtaDCS_1} shows the $\gamma p\to p\eta$ differential cross sections 
for the two different $\eta$ decay modes (\ref{ReactionEta1}) and (\ref{ReactionEta2}). 
We have excluded those data points in the analysis showing a Monte-Carlo acceptance 
of \text{$< 5\,\%$}. The data sets show excellent agreement. 

We have checked the consistency of the two cross-section measurements by forming the 
ratio of partial widths of the two $\eta$~decay modes. Since the cross sections
in Fig.~\ref{Figure:EtaDCS_1} are corrected for the decay branching ratios, we
determine
\begin{equation}
\frac{\Gamma_{\eta\rightarrow 3\pi^0}}{\Gamma_{\eta\rightarrow
2\gamma}}=\frac{\biggl[{\rm
\frac{d\sigma}{d\Omega}}(E_{\gamma},\cos\theta_{\rm{cm}})\biggr]_{\eta\rightarrow
3\pi^0}\biggl[\frac{\Gamma_{\eta\rightarrow
3\pi^0}}{\Gamma_{total}}\biggr]_{\rm{PDG}}} {\biggl[{\rm
\frac{d\sigma}{d\Omega}}(E_{\gamma},\cos\theta_{\rm{cm}})\biggr]_{\eta\rightarrow
2\gamma}\biggl[\frac{\Gamma_{\eta\rightarrow
2\gamma}}{\Gamma_{total}}\biggr]_{\rm{PDG}}}
\end{equation}

\noindent
for each ($E_\gamma$,\,cos\,$\theta_{cm}$) bin. The values are weighted with
their squared inverse errors and histogrammed (Fig. \ref{ratioDecayModes}).
We derive a peak position of
\begin{equation}
  \frac{\Gamma_{\eta\rightarrow 3\pi^0}}
       {\Gamma_{\eta\rightarrow 2\gamma}} = (0.8133\pm 0.0006\,_{\rm stat.}\pm 0.0138\,_{\rm sys.})\,,
\end{equation}
\noindent
where the systematic error is derived from considering all systematic
uncertainties discussed in Section~\ref{Subsection:SystematicUncertainties}. 
We do not claim a new measurement of $\eta$ branching fractions here, but
have rather used this number to check our reconstruction efficiency. The 
Particle Data Group gives two values for this branching ratio~\cite{Amsler:2008zz}: 
$(0.829\pm 0.007)$ is the mean value of all direct measurements while a combined 
fit to all partial decay widths yields a value of $(0.828\pm 0.006)$. Both
PDG values are consistent with our result.

\subsection{Differential Cross Sections $\boldsymbol{d\sigma/d\Omega}$ for
$\boldsymbol{\gamma p\to p\eta}$ -- Combined Data Set --}
\label{Subsection:CrossSectionsEtaCombined}

Since the differential cross sections for the two $\eta$ decay modes are
consistent, we have calculated error-weighted mean values. These are presented 
in Fig.~\ref{Figure:EtaDCS_2} as functions of energy and the $\eta$ production
angle. Since the cross sections change rather smoothly, only few resonances
are likely to contribute to the process. The $N(1535)S_{11}$ state is known
to dominate the threshold region resulting in a flat distribution
in cos\,$\theta_{\rm cm}$; interference with other amplitudes leads to
deviations from that flat distribution. At higher energies, above $E_\gamma
= 1.5$~GeV, the development of a forward peak indicates important contributions
from $t$-channel $\rho$ and $\omega$ exchange.

Fig.~\ref{Figure:EtaDCS_2} also shows a comparison of our new $\eta$ results
to published results from CB-ELSA~\cite{Crede:2003ax}. The agreement between the
two data sets is very good at lower energies. However, the differential cross
sections reported by CB-ELSA show somewhat larger discrepancies at higher energies 
and forward angles. Above $E_\gamma = 2.5$~GeV and in the forward most angle bins,
CB-ELSA results are approximately 30\,\% larger than our new CBELSA/TAPS
results at these energies, but are still consistent within the errors. We believe 
that the discrepancy is due to underestimated background in the CB-ELSA data in 
the low-statistics forward most bins where fewer than 10~events were observed.

\subsection{Differential Cross Sections $\boldsymbol{d\sigma/d\Omega}$ for
$\boldsymbol{\gamma p\to p\eta\,^\prime}$ at an Electron-Beam Energy of
$\boldsymbol{E_{{\rm e}^-} = 3.18}$~GeV}
\label{Subsection:CrossSectionsEtap}

\begin{figure*}[t]
  \epsfig{file=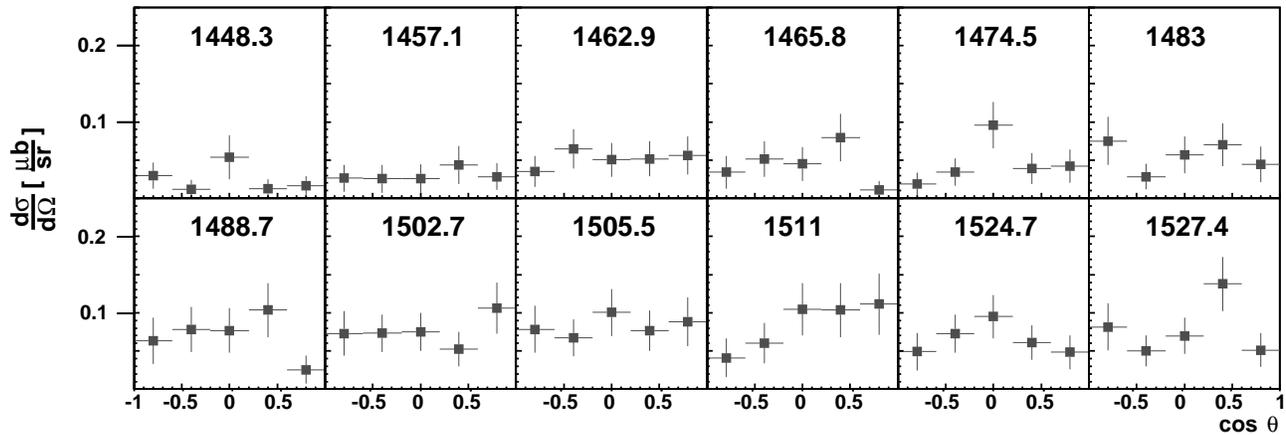,width=1.0\textwidth}
  \caption{\label{Figure:EtapWQThreshold}Differential cross sections for the
    reaction $\gamma p\to p\eta\,^\prime\to 2\pi^0\eta\to p6\gamma$ close to
    the reaction threshold ({\tiny\color{gray07}$\blacksquare$}) determined 
    for individual tagger channels using 5 cos\,$\theta_{\rm c.m.}^{\,\eta\,
    ^\prime}$ bins of width 0.4. Energies in the plots are given in MeV. Though
    limited in statistics, all angular distributions appear to be flat indicating 
    s-wave behavior of the reaction at the threshold.}
\end{figure*}

Fig.~\ref{Figure:EtapDCS} shows the differential cross sections for
the reaction $\gamma p\to p\eta\,^\prime\to 2\pi^0\eta$ (reaction 
\ref{ReactionEtap}) using cos\,$\theta_{\rm c.m.}^{\,\eta\,^\prime}$ 
bins of width 0.2. The data cover the full angular range. Very similar 
to $\eta$ photoproduction, a rather flat angular distribution is observed at 
low energies suggesting $s$-channel resonance production near threshold. 
Both data sets also show a continuing increase in slope at forward angles, 
which becomes more prominent at higher energies. This forward peak is most 
likely due to $t$-channel exchange mechanisms. Moreover, our new data
indicate a decrease in the forward most bin, which has not been observed
before. Above 2~GeV in photon energy, growth at backward angles could be 
indicative of $u$-channel contributions. The overall agreement between 
the CBELSA/TAPS and CLAS data is good at threshold to fair above 
$E_\gamma = 1800$~MeV.

\begin{figure}[b]
  \epsfig{file=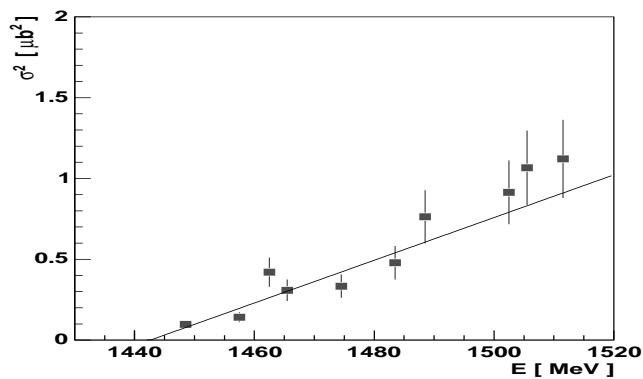,width=0.50\textwidth,height=0.3\textwidth}
  \caption{\label{Figure:EtapThreshold} Shown is the linear energy dependence of
    the squared total cross section, $\sigma_{\rm tot}^2$\,, for the reaction 
    $\gamma p\to p\eta\,^\prime$ close to the reaction threshold of 
    $E_{\rm thres}\approx 1447$~MeV.}
\end{figure}

The s-wave behavior of the reaction close to the reaction threshold is 
apparent from the experimental data. Fig.~\ref{Figure:EtapWQThreshold} 
shows the differential cross sections for $\gamma p\to p\eta\,^\prime$ at 
and close to the reaction threshold using 5~cos\,$\theta_{\rm c.m.}^{\,\eta\,
^\prime}$~bins of width~0.4. The cross sections have been determined for
individual fibers of the tagging system and cover the full angular range.
The data points suffer from low statistics, but are overall consistent with
flat angular distributions. The expected energy dependence of the reaction 
at the threshold is given by~\cite{Krusche:2003ik}:
\begin{equation}
  \label{Equation:Threshold}
  \sigma(E_\gamma)\propto (E_\gamma - E_{\rm thres})^{l+1/2}\qquad 
l=0,\,1,\,2,\,...\,,
\end{equation}
\noindent
where $l$ denotes $s,\,p,\,d\,...$ waves, etc. For $s$-wave dominance, a
linear energy dependence of the squared total cross section is thus expected 
close to the reaction threshold. The differential cross sections shown in 
Fig.~\ref{Figure:EtapWQThreshold} have been used to determine the total
$\eta\,^\prime$ cross section and to study the energy dependence. Since the 
data cover the full angular range, no extrapolation is needed.
Fig.~\ref{Figure:EtapThreshold} shows the incoming photon energy plotted 
versus the squared total cross section. A linear dependence is clearly 
observed and a fit determines the energy threshold to $(1442.6\pm 3.8)$~MeV,
which is compatible with the value of $(1446.38\pm 0.48)$~MeV derived from 
the $\eta\,^\prime$ mass listed in the PDG~\cite{Amsler:2008zz}. The mass of 
the $\eta$~meson was determined by the TAPS/A2~Collaboration in a very similar 
procedure~\cite{Krusche:1995cr,Krusche:1995nv}.

\subsection{\label{sigmatot}The total cross section}
\begin{figure}[pb]
  \epsfig{file=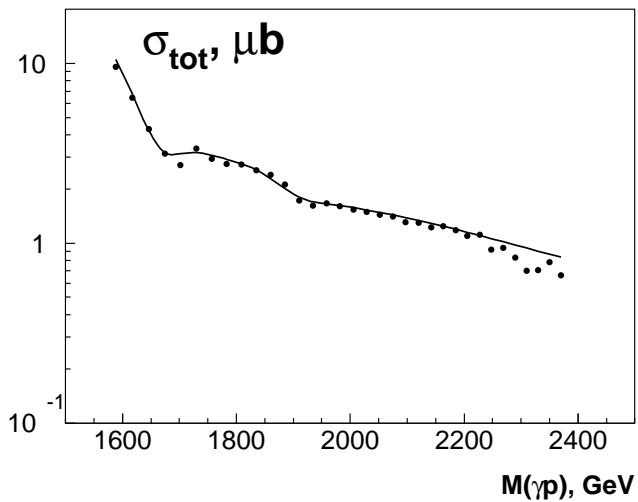,width=0.48\textwidth}
  \vspace{-4mm}
  \caption{\label{Fig:P11P13}Total $\gamma p\to p\eta$ 
  cross section. The data points ({\color{gray07}$\bullet$}) are 
  calculated by summation of the differential cross section, the grey 
  line represents the result of our previous partial wave analysis.} 
\end{figure}

Fig.~\ref{Fig:P11P13} shows the total cross section for $\eta$
photoproduction. Due to the complete solid angle coverage, no
extrapolation is required and the data points are truly experimental.
In the low-energy range, the $S_{11}$ partial wave dominates the cross 
section. The solid line represents our previous PWA solution and is not 
a fit to these data; the two states $N(1720)P_{13}$ and $N(2070)D_{15}$ 
saturated the total cross section~\cite{Crede:2003ax}. A new PWA solution 
including the new CBELSA/TAPS data presented here and other data sets is 
in preparation. 
It is clear that single- and double-polarization variables are required
to firmly establish resonance contributions. Coupled channel fits to many 
reactions can also help; in particular, when three-body final states are 
included, the phase of two-particle partial-wave amplitudes is tested in 
the crossed channel.

\begin{figure}[t]
  \vspace{-4mm}
  \epsfig{file=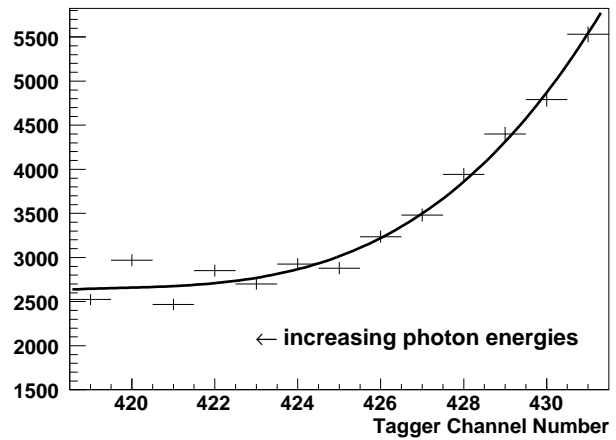,width=0.50\textwidth}
  \caption{\label{Fig:yields}Total $\eta\to\gamma\gamma$ yields per 
  tagger channel number covering the energy range of the anomaly observed 
  in the total $\eta$ cross section. The data point at 1.73~GeV/$c^2$ 
  (see Fig.~\ref{Fig:P11P13}) is based on channels 421-426. No statistically 
  significant enhancement can be seen in the excitation function.}
\end{figure}

A small anomaly is observed at 1.73~GeV/$c^2$ in the total $\eta$ cross 
section. As discussed earlier, recent data off the neutron show a pronounced 
bump-like structure at 1.68~GeV~\cite{Jaegle:2008ux}, which has been
suggested to signal the existence of a narrow baryon state with 
$(M\approx~1.68~, \Gamma\leq~30)$~MeV/$c^2$. In particular, the possibility
that this state could be the $N(1650)P_{11}$, non-strange member of an 
anti-decuplet of pentaquarks is certainly interesting~\cite{Diakonov:1997mm,
Polyakov:2003dx,Diakonov:2003jj}. Fig.~\ref{Fig:yields} shows the total 
number of $\eta\to\gamma\gamma$ events for the tagger channels 419-431. The 
data have been fitted using a polynomial to indicate the smooth behavior of 
the distribution. The data point at 1.73~GeV/$c^2$ in the total cross section 
is based on the photon energy interval $E_{\gamma}\,\in\,[1100,\,1150]$~MeV 
defined by the tagger channels $E_{\gamma}\,\in\,[421,\,426]$. No statistically 
significant enhancement is observed in Fig.~\ref{Fig:yields} over the small 
energy range under investigation to explain the anomaly and a narrow state 
compatible with the observation in the total cross section can be ruled out. 
We believe that this structure is an instrumental effect originating from 
tagger cluster-size corrections of low-rate Tagger-Or-data used in the 
photon-flux determination, which occurs only for those channels.

\begin{figure}[b]
  \vspace{0.5mm}
  \epsfig{file=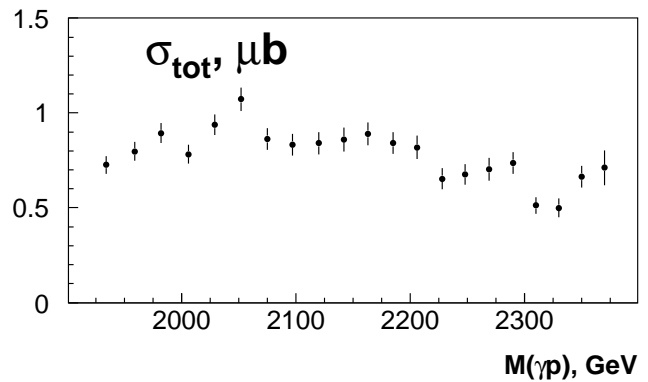,width=0.48\textwidth}
  \vspace{-5mm}
  \caption{\label{Fig:etaptot}Total $\gamma p\to p\eta\,^{\prime}$ cross
  section. The data points ({\color{gray07}$\bullet$}) are calculated by 
  summation of the differential cross section.}
\end{figure}

In Fig. \ref{Fig:etaptot}, the total cross section for $\eta\,^\prime$
photoproduction is displayed. Again, due to the complete solid angle coverage, 
no extrapolation is required. At 2~GeV in the invariant mass and above, the 
cross section for $\eta\,^\prime$ production is about 50\,\% of that for $\eta$ 
production assuming that the processes are dominated by $\rho,~\omega$-exchange
resonances. For a pseudoscalar mixing angle of $\Theta_{\rm PS}=-19.3^\circ$, the 
non-$s\bar{s}$ components of $\eta\,^\prime$ and $\eta$ differ by a factor 1/2. 
The similarity of the two numbers suggests that the dynamics of $\eta$ and 
$\eta\,^\prime$ photoproduction is similar. 

\section{Recent Results of the Bonn-Gatchina Model \label{Section:PWA}}
The analysis of previous CB-ELSA data on photoproduction of $\eta$-mesons
\cite{Crede:2003ax,Bartholomy:2007zz} revealed two surprises:

(1) The analysis suggested a new resonance with spin and parity
$J^{P}=5/2^{-}$, $N(2070)D_{15}$ \cite{Anisovich:2005tf}. The total
cross section was nearly saturated with three resonances, the
well-known $N(1535)S_{11}$, the $N(1720)P_{13}$, and the new
resonance.

(2) The strong coupling of $N(1720)P_{13}\to N\eta$ was also
unexpected; in MAID \cite{Chiang:2002vq,Tiator:2006he}, the
$N(1710)P_{11}\to N\eta$ was very significant while $N(1720)P_{13}$
hardly contributed to $\eta$ photoproduction.

The pattern of states contributing most to $\eta$ photoproduction,
$N(1535)S_{11}$, $N(1720)P_{13}$, $N(2070)D_{15}$, was interpreted
as a sequence of quark model states with total intrinsic angular
momenta $L=1,2,3$ and $S=1/2$ coupling to $J^P=1/2^-,3/2^-,5/2^-$.
The regularity of this pattern was used to argue that the seed of
$N(1535)S_{11}$ should be of a three-quark nature; due to the presence
of $S$-wave thresholds, the state may attract large $N\eta$ and
$\Sigma\pi$ molecular components.

The relative strength of the two nucleon excitations at an incoming 
photon energy of 1700~MeV in $\gamma p\to p\eta$ remains disputed. The 
Gie\ss en group~\cite{Shklyar:2006xw} found - like MAID -  
$N(1710)P_{11}\to N\eta$ to provide a significant contribution. Other 
coupled channel analyses confirmed the dominance of $N(1720)P_{13}$ 
(relative to $N(1710)P_{11}$). In~\cite{Shyam:2008fr} and~\cite{Nakayama:2008tg}, 
a large variety of $\eta$ production data was fitted using an effective 
Lagrangian approach; in both analyses, the $N(1720)P_{13}$ contribution 
was considerably larger than that of $N(1710)P_{11}$. A chiral quark model 
approach complemented with a one-gluon exchange model~\cite{He:2008ty} 
arrived at the same conclusion.

Restricted to photoproduction data, the best solution for the new data 
presented here (in terms of its $\chi^2$ value) still supports the 
dominance of the three nucleon resonances, $N(1535)S_{11}$, $N(1720)P_{13}$,
and $N(2070)D_{15}$ in $\eta$ photoproduction. However, this solution
is incompatible with data on $\pi^- p\to n\eta$. This is presently 
investigated further and will be subject of a forthcoming publication 
of the Bonn-Gatchina partial wave analysis group.

\section{\label{Section:Summary}Summary}
In summary, we have presented data on the photo-produced $\eta$ and
$\eta\,^\prime$ cross sections from the reactions $\gamma p\to p\eta$ 
with $\eta\to 3\pi^0\to 6\gamma$ as well as $\eta\to 2\gamma$ and from 
the reaction $\gamma p\to p\eta\,^\prime$ with $\eta\,^\prime\to\pi^0
\pi^0\eta\to 6\gamma$. The continuous beam from the ELSA accelerator and
the fiber detector of the tagging system provided tagged-photons in the 
energy range from 850 to 2550~MeV. The results are in very good agreement 
with previous measurements, but extend over the full angular range in 
$\rm cos\,\theta_{cm}$ of the $\eta$ and $\eta\,^\prime$~meson. The inclusion 
of the new $\eta$ data into a multi-channel partial wave analysis is in 
preparation. The threshold behavior of the $\eta\,^\prime$~data indicate 
$s$-wave dominance.

\subsection*{Acknowledgments}
We thank the technical staff at ELSA and at all the participating
institutions for their invaluable contributions to the success of
the experiment. We acknowledge financial support from the National 
Science Foundation (NSF), Deutsche Forschungsgemeinschaft (DFG) 
within the SFB/TR16 and from Schweizerischer Nationalfond. The collaboration 
with St. Petersburg received funds from DFG and the Russian Foundation 
for Basic Research.

\end{document}